\documentclass[aps,prd,a4paper,twocolumn,10pt]{revtex4-1}
\usepackage[colorlinks,linkcolor=blue,anchorcolor=blue,citecolor=blue,urlcolor=blue]{hyperref}
\usepackage{amsmath,amsfonts,amssymb}
\usepackage{color,graphicx,epsfig,subfigure}
\usepackage{diagbox,txfonts,acronym,indentfirst,booktabs}

\allowdisplaybreaks
\newcommand{\bea}{\begin{eqnarray}}
\newcommand{\eea}{\end{eqnarray}}
\newcommand{\balign}{\begin{aligned}}
\newcommand{\ealign}{\end{aligned}}

\newcommand{\TRC}{MOE Key Laboratory of TianQin Mission, TianQin Research Center for
Gravitational Physics \& School of Physics and Astronomy, Frontiers
Science Center for TianQin, Gravitational Wave Research Center of
CNSA, Sun Yat-sen University (Zhuhai Campus), Zhuhai 519082, China.}
\begin{document}
\title{Bayesian analysis of gravitational wave memory effect with TianQin}

\author{Shuo Sun}
\author{Changfu Shi}
\email{Corresponding author. Email: shichf6@mail.sysu.edu.cn}
\author{Jian-dong Zhang}
\author{Jianwei Mei}

\affiliation{\TRC}
\newacro{GR}{general relativity}
\newacro{GW}{gravitational wave}
\newacro{MBHB}{massive black hole binary}
\newacro{BH}{black hole}
\newacro{ASD}{amplitude spectral density}
\newacro{PSD}{power spectral density}
\newacro{BMS}{Bondi-Metzner-Sachs}
\newacro{SNR}{signal-to-noise ratio}
\newacro{PN}{post-Newtonian}
\newacro{CCE}{Cauchy-characteristic extraction}
\newacro{IMR}{Inspiral-Merger-Ringdown}
\date{\today}
\begin{abstract}

The memory effect in gravitational waves is a direct prediction of general relativity. The presence of the memory effect in gravitational wave signals not only serves as a test for general relativity but also establishes connections between soft theorem, and asymptotic symmetries, serving as a bridge for exploring fundamental physics. Furthermore, with the ongoing progress in space-based gravitational wave detection projects, the gravitational wave memory effect generated by the merger of massive binary black hole binaries is becoming increasingly significant and cannot be ignored. In this work, we perform the full Bayesian analysis of the gravitational wave memory effect with TianQin. The results indicate that the memory effect has a certain impact on parameter estimation but does not deviate beyond the 1$\sigma$ range. Additionally, the Bayes factor analysis suggests that when the signal-to-noise ratio of the memory effect in TianQin is approximately 2.36, the $\text{log}_{10}$ Bayes factor reaches 8. This result is consistent with the findings obtained from a previous mismatch threshold.
\end{abstract}
\maketitle

\section{Introduction}

The ground-based GW detectors LIGO and Virgo have observed numerous gravitational wave (GW) events generated by compact binary coalescences (CBCs) since 2015 \cite{LIGOScientific:2018mvr,LIGOScientific:2020ibl,LIGOScientific:2021djp}. This discovery has introduced a novel approach to test general relativity \cite{LIGOScientific:2019fpa,LIGOScientific:2020tif,LIGOScientific:2021sio} and study the nature of black holes \cite{LIGOScientific:2018jsj,LIGOScientific:2020kqk,LIGOScientific:2021psn,Isi:2019aib}. In the near future, more ground-based gravitational wave detectors, i.e., KAGRA \cite{Aso:2013eba}, Einstein telescope \cite{Punturo:2010zz}, and Cosmic explorer \cite{Reitze:2019iox}, will join efforts to detect GWs. This will increase the precision of GW detection, expanding the observed mass range and distance of compact binary systems. It will also contribute to a deeper understanding of the nature of gravity and black holes.

Due to the presence of terrestrial noise, ground-based GW detectors can only detect high-frequency GW signals from the merger of compact binary systems with a total mass in the range of a few hundred solar masses ($\text{M}_{\odot}$). The space-borne GW detection missions have been proposed to overcome this limitation and enable the detection of much lower frequency GW signals. The space-borne GW detectors, TianQin \cite{TianQin:2015yph,TianQin:2020hid}, LISA \cite{LISA:2017pwj}, and Taiji \cite{Hu:2017mde}, are scheduled to launch and operate in the 2030s. The sensitivity range of TianQin for detecting GW signals is from $10^{-4}$ Hz to $1$ Hz. Unlike LISA and Taiji, TianQin consists of three satellites orbiting the Earth, which are arranged in a constellation of approximately equilateral triangles with arm lengths of about $3 \times 10^5$ km. TianQin aims to detect the GW signals from the merger of massive black hole binaries (MBHBs), as well as the inspiral of stellar mass black hole binaries (SBHBs) \cite{Liu:2020eko,Liu:2021yoy}, extreme-mass-ratio inspirals (EMRIs) \cite{Fan:2020zhy}, stochastic gravitational wave background (SGWB) \cite{Liang:2021bde,Cheng:2022vct}, and galactic double white dwarf binaries (DWDs) \cite{Huang:2020rjf}.

TianQin is capable of detecting the GW signals from MBHBs, providing the potential to constrain the cosmological parameters \cite{Zhu:2021aat},  measure the Hubble constant by gravitational lensing \cite{Lin:2023ccz,Huang:2023prq} and test general relativity \cite{Shi:2019hqa,Bao:2019kgt,Zi:2021pdp,Shi:2022qno}. Furthermore, the potential for high signal-to-noise ratios (SNRs) of signals of MBHBs on TianQin, opens up the possibility to study the strong-field gravitational effects. The memory effect \cite{zel1974,throne1987,PhysRevLett.67.1486,PhysRevD.46.4304} is one such strong-field effect, associated with the energy and angular momentum flux released during black hole mergers \cite{Flanagan:2015pxa}, resulting in a permanent spacetime change. 

Detecting the memory effect is essential, both from a theoretical aspect and for precise parameter estimation. From a theoretical perspective, the detection of the memory effect can provide a way to test asymptotic symmetries and soft gravitons \cite{Strominger:2014pwa} (referred to as the infrared triangle \cite{Strominger:2017zoo}), as well as providing an alternative approach to testing general relativity \cite{Du:2016hww, Seraj:2021qja,Tahura:2021hbk,Hou:2021oxe,Hou:2021bxz,Hou:2023pfz}. From a parameter estimation perspective, considering the memory effect leads to more precise parameter estimation results and also provides a means of verifying the accuracy of the waveform \cite{Ashtekar:2019viz}.

The idea of detecting the GW memory effect through LIGO was first proposed in the 1980s \cite{throne1987}. In recent years, many studies have discussed detecting the memory effect through Pulsar Timing Arrays (PTAs) \cite{10.1111/j.1745-3933.2009.00758.x,vanHaasteren:2009fy,Pshirkov:2009ak,Cordes:2012zz,madison,NANOGrav:2015xuc}, advanced LIGO and Virgo \cite{Lasky:2016knh,McNeill:2017uvq,Divakarla:2019zjj,Boersma:2020gxx} and searching for the memory effect in existing GW data \cite{Hubner:2019sly,Khera:2020mcz,Hubner:2021amk,Zhao:2021hmx}. These results indicate that it is currently challenging to detect the memory effect using current ground-based detectors. Furthermore, the 12.5-year data from NANOGrav has not provided convincing evidence for the existence of a memory effect signal \cite{Agazie:2023eig}. Additionally, some studies have investigated the possibility of detecting the memory effect through space-based GW detectors \cite{Islo:2019qht,Sun:2022pvh}. The findings of these studies suggest that space-based GW detectors may have the potential to directly detect the GW memory effect produced by MBHBs. 

Previous studies have indicated that the memory effect exhibits a relatively high SNR on space-based GW detectors \cite{Islo:2019qht,Sun:2022pvh}. However, since most current GW waveforms do not include the memory effect, investigating whether the memory effect affects parameter estimation remains meaningful. The Bayesian inference is a commonly used method for parameter estimation in GW astronomy. To ensure the accuracy of parameter estimation results and computational efficiency, it is essential to employ a method that is both relatively accurate and computationally efficient for calculating memory effect waveforms. Moreover, due to the current necessity of computing memory effect waveforms in the time domain and the complexity of the response for space-based gravitational wave detectors, there are challenges in simultaneously considering waveform computation and the implementation of the TianQin response in signals.

In this work, our focus is on developing a method for performing Bayesian inference on the memory effect detected by TianQin using simulated data from MBHBs. This method involves utilizing Bondi-Metzner-Sachs (BMS) balance laws \cite{Flanagan:2015pxa, Mitman:2020pbt} to compute memory effect waveforms and applying Time Delay Interferometry (TDI) response to these waveforms for TianQin. This enables us to calculate the posterior distribution of parameters to investigate the impact of memory effect on parameter estimation results. We will study the detection SNR threshold for memory effect on TianQin by calculating the Bayes factor obtained from the comparison of waveforms with and without memory effect in parameter estimation. Additionally, it is compared with the SNR derived from the threshold of the mismatch between waveforms with and without memory effects, as obtained in previous work \cite{Sun:2022pvh}.  

The paper is organized as following. In Sec. \ref{BMSBL} we give a brief introduction to the memory and the main method for calculating the displacement memory. In Sec. \ref{sec3}, we introduce the time domain TDI response for TianQin and recall the basic theorem of Bayes inference.  In Sec. \ref{sec4}, we presented simulated data generated by our code. In Sec. \ref{sec5}, we will present our main findings regarding parameter estimation, including results for Bayes factor and the comparison of Bayes factor with the mismatch threshold. Finally, we present the summary and discussion in Sec. \ref{conclusion}. Throughout the work, we use $G=c=1$ unless otherwise stated.

\section{The gravitational wave memory}\label{BMSBL}
The gravitational wave memory effects have been extensively studied over the past few decades \cite{zel1974,throne1987,PhysRevLett.67.1486,PhysRevD.46.4304,Thorne:1992sdb,Favata:2008ti,Favata:2009ii,Favata:2008yd,Favata:2010zu,Favata:2011qi}. In recent years, it has been recognized that there is a connection between memory effects and asymptotic symmetries \cite{Strominger:2014pwa}. Due to different BMS transformations \cite{Bondi:1962px,Sachs:1962wk,Barnich:2009se}, there are two new memory effects, namely the spin memory effect corresponding to superrotation \cite{Pasterski:2015tva}, and the center-of-mass memory effect corresponding to superboost \cite{Flanagan:2015pxa,Nichols:2018qac,Mitman:2020pbt}. The memory effect that we commonly refer to, corresponding to supertranslation, is called the displacement memory effect, based on the physical phenomenon it induces. All three memory effects are categorized into linear and nonlinear components based on their distinct origins. It has been suggested to rename linear and nonlinear memory effects as ordinary and null memory effects, respectively, to better represent their true nature \cite{Bieri:2013ada}.

The methods for computing the waveform of the memories have been extensively studied in the past few years. One way to compute displacement memory is to use the post-Newtonian (PN) approximation and various post-processing techniques \cite{Favata:2008yd,Favata:2009ii,Favata:2010zu,Talbot:2018sgr}. Another post-processing method to calculate memory effects is related to BMS transformations, where memory effects can be computed through variations in BMS charges and their associated fluxes, known as BMS flux balance laws \cite{Flanagan:2015pxa, Mitman:2020pbt}. The results of numerical relativity have verified the accuracy of the memory waveforms calculated using this method \cite{Mitman:2020bjf}.

Due to the very low SNR generated by the spin memory effect on detectors, its detection is currently unlikely \cite{Sun:2022pvh}, and this also applies to the center-of-mass memory effect \cite{Nichols:2018qac}. Therefore, in this study, we focus only on the displacement memory effect. For the sake of computational accuracy and efficiency, we utilize BMS balance laws to calculate the waveforms of the displacement memory. 

The method of BMS flux balance laws is mainly based on \cite{Mitman:2020pbt}. We will provide a brief summary of the methods used to calculate memory as follows. 

The GW strain, which we denote as $h$, can be represented in spherical harmonic basis,
\bea h(u)=\sum_{\ell\geq2}\sum_{|m|\leq\ell}h_{\ell m}(u) ~_{-2}Y_{\ell m}(\iota,\varphi)\,,\label{gwstrain}
\eea
where $_{-2}Y_{\ell m}(\iota,\varphi)$ is spin-weighted $-2$ spherical harmonics, $\iota$ is the inclination angle, $\varphi$ is the reference phase of source and $u=t-r$ is the Bondi time. The displacement memory, which we denote as $h_{\rm mem}$, can be calculated as 
\bea
h_{\rm mem}=\frac{1}{2}\bar{\eth}^{2}\mathfrak{D}^{-1}\left[\Delta m+\frac{1}{4}\int^{u}_{-\infty}|\dot{h}|^{2}du\right]\,,\label{dis}
\eea 
where $\dot{h}$ is Bondi time derivative of $h$. The operators $\eth$ and $\bar{\eth}$ in the Newman-Penrose convention are defined as \cite{roger1},
\bea
\balign
&&\eth_{s}Y_{\ell m}=+\sqrt{(\ell-s)(\ell+s+1)}_{s+1}Y_{\ell m}\,,\\
&&\bar{\eth}_{s}Y_{\ell m}=-\sqrt{(\ell+s)(\ell-s+1)}_{s-1}Y_{\ell m}\,.
\ealign
\eea
Applying the operators $D^{2}=\bar{\eth}\eth$ and $\mathfrak{D}=\frac{1}{8}D^{2}(D^{2}+2)$ to the spin weight 0 spherical harmonics $Y_{lm}$ \cite{Mitman:2020pbt},
\bea 
\balign
D^{2}Y_{\ell m}&=-\ell(\ell+1)Y_{\ell m}\,,\\
\mathfrak{D}Y_{\ell m}&=\frac{1}{8}(\ell+2)(\ell+1)\ell(\ell-1)Y_{\ell m}\,.\label{eq4b}
\ealign
\eea
The operator $\mathfrak{D}^{-1}$ is the inverse of $\mathfrak{D}$ and is defined by projecting out the $\ell \leq 1$ mode of spherical harmonics $Y_{lm}$ \cite{Mitman:2020pbt},
\bea\mathfrak{D}^{-1}Y_{\ell m}=\left\{\begin{matrix}0&:&\ell\leq 1\,,\cr
\left[\frac{1}{8}(\ell +2)(\ell+1)\ell(\ell-1)\right]^{-1}Y_{\ell m}&:&\ell\geq2\,.\end{matrix}\right.\label{mkD}\eea
The $\Delta m = m(u)-m(-\infty)$ in the right-hand side of Eq. (\ref{dis}) is the change in BMS charge, that is, the Bondi mass aspect, and it is the source of the linear (or ordinary) part of displacement memory. The second term is the nonlinear (or null) part and it comes from the energy flux carried away by GWs. For the linear part, it can be written in terms of Weyl scalar $\Psi_{2}$ and $h$ as 
\bea m=-{\rm Re}\left[\Psi_2+\frac{1}{4}\dot{h}\bar{h}\right]\,,\eea
where Re means the real part of the function.

In the BBH merger, the contribution of $\Delta m$ to the memory effect is extremely small and can be neglected \cite{Mitman:2020pbt,Mitman:2020bjf,Ashtekar:2019viz,Ashtekar:2019rpv}. In this work, we only consider the nonlinear part of the displacement memory, and we will refer to it simply as ``memory" in the subsequent sections.
\section{Methodology}\label{sec3}
\subsection{Detector response}
  Due to the changes in arm length caused by spacecraft motion, Time Delay Interferometry (TDI) technique has been proposed to cancel laser frequency noise \cite{Armstrong_1999,Tinto:1999yr,Estabrook:2000ef,Dhurandhar:2001tct,Tinto:2003vj,Vallisneri:2004bn}. We use the uncorrelated combinations A, E, and T combinations \cite{Prince:2002hp} for the detector response. We adopt the TianQin orbit proposed in \cite{Hu:2018yqb} and the calculation of TDI is based mainly on the method of \cite{Katz:2022yqe}. 
\begin{figure}[t]
\centering
\includegraphics[scale=0.35]{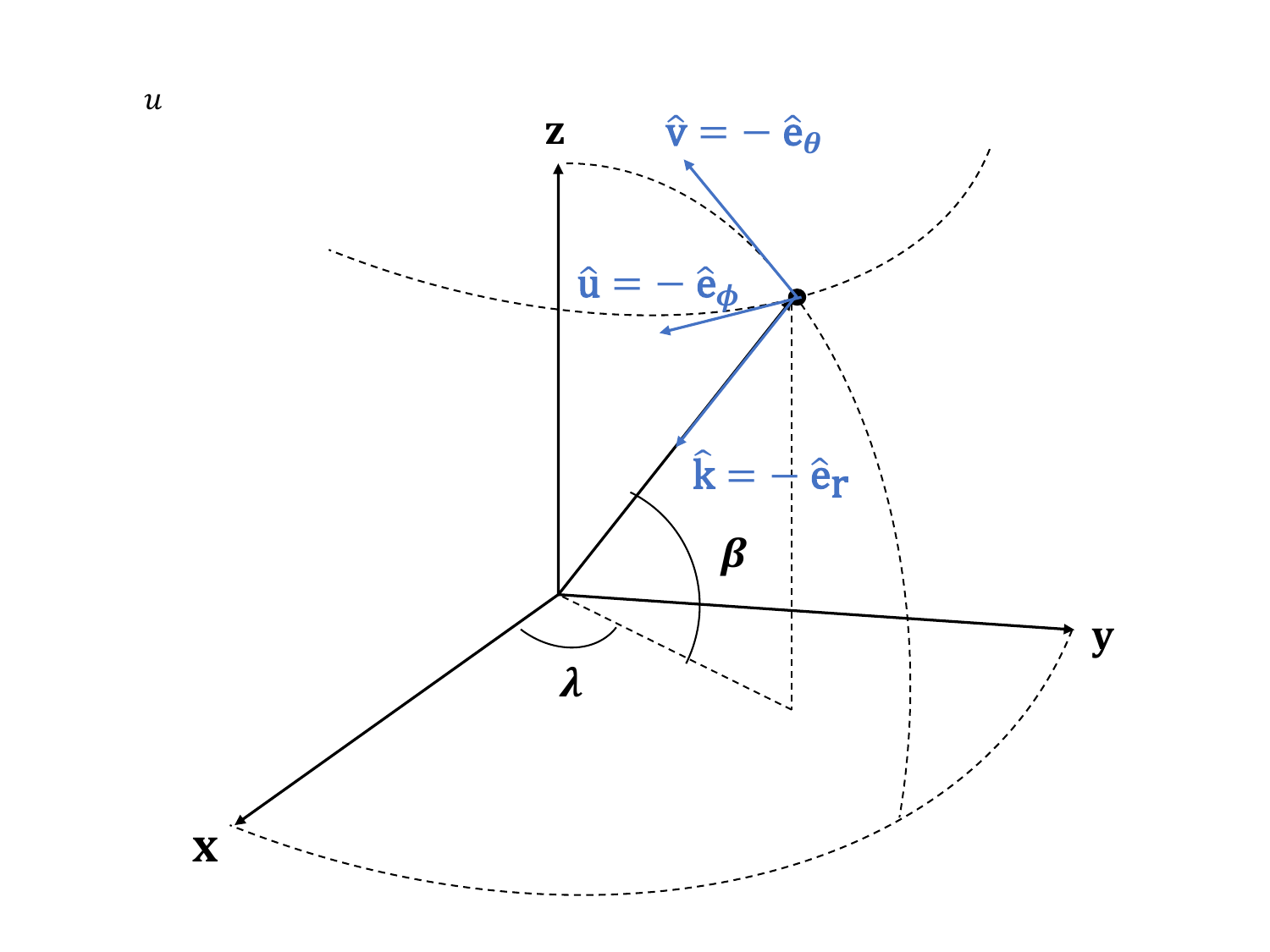}
\caption{Orientation of the source in the SSB coordinate system. The GW propagation vector is $\hat{\mathbf{k}}$, and the polarization vectors are $\hat{\mathbf{u}}$ and $\hat{\mathbf{v}}$.}
\label{ssb}
\end{figure}

To calculate the A, E, and T channels, it is necessary to first transform the GW signal from the source coordinate system to the detector coordinate system. The TianQin orbit and the sky location of GW sources can be described by the Cartesian coordinate system of the Solar System Barycenter (SSB), defined as $(\boldsymbol{x}, \boldsymbol{y}, \boldsymbol{z})$, and can be represented in the spherical coordinate system $(\theta, \phi)$ for convenience. The basis vectors can be constructed by the ecliptic latitude $\beta=\pi/2-\theta$ and the ecliptic longitude $\lambda=\phi$ \cite{Katz:2022yqe},
\bea
\balign
\hat{\mathbf{e}}_{r}&=(\cos \beta \cos\lambda,\,\cos\beta\sin\lambda,\,\sin\beta ),\\ 
\hat{\mathbf{e}}_{\theta}&=(\sin \beta \cos\lambda,\,\sin\beta\sin\lambda,\,-\cos\beta ),\\
\hat{\mathbf{e}}_{\phi}&=(-\sin\lambda,\cos\lambda,0).
\ealign
\eea
The propagation vector can be defined as
\bea
\balign
\hat{\mathbf{u}}&=-\hat{\mathbf{e}}_{\phi},\\
\hat{\mathbf{v}}&=-\hat{\mathbf{e}}_{\theta},\\
\hat{\mathbf{k}}&=-\hat{\mathbf{e}}_{r},
\ealign
\eea
which can produce a direct orthonormal basis in $(\hat{\mathbf{u}},\hat{\mathbf{v}},\hat{\mathbf{k}})$. The SSB spherical coordinates ($\theta,\phi$) as illustrated in Fig. \ref{ssb}.

The GW signals in SSB frame can be described by
\bea
\balign
h_{+}^{\text{SSB}} &= h_{+}^{\text{S}}\cos 2\psi - h_{\times}^{\text{S}}\sin 2\psi,\\
h_{\times}^{\text{SSB}} &= h_{+}^{\text{S}}\sin 2\psi + h_{\times}^{\text{S}}\cos 2\psi.
\ealign
\eea
where $h_{+}^{\text{S}}$ and $h_{\times}^{\text{S}}$ are the real and imaginary part of Eq. (\ref{gwstrain}) in the source frame, which we take the form of $h^{\text{S}}$ as $h^{\text{S}} = h_{+}^{\text{S}} - ih_{\times}^{\text{S}}$.

\begin{figure}[t]
\centering
\includegraphics[scale=0.32]{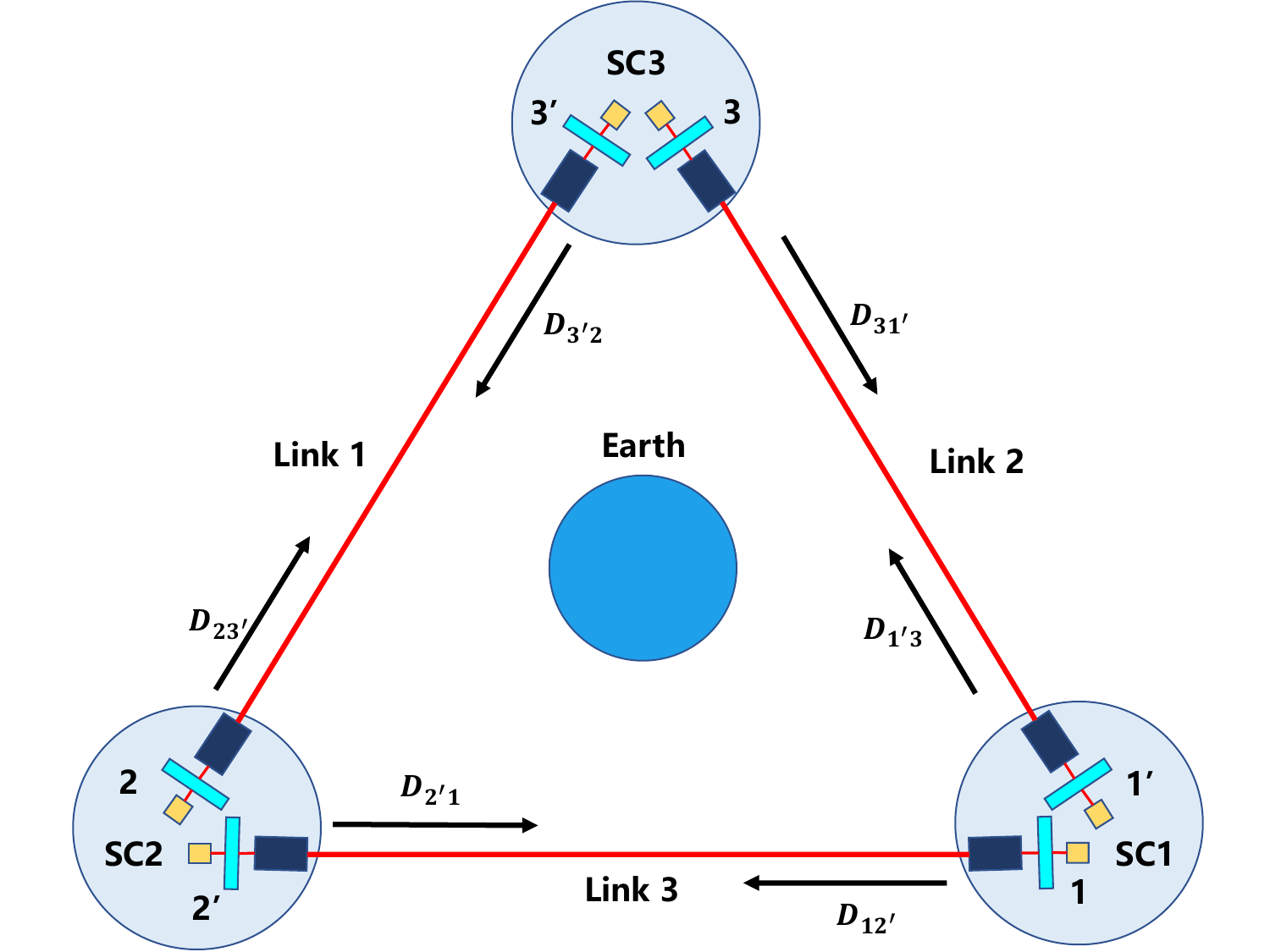}
\caption{Labeling conventions used in TianQin split interferometer.}
\label{tqtdi}
\end{figure}

The TianQin constellation is shown in Fig. \ref{tqtdi}. The three spacecrafts are labeled as ``SC$_{i}$" where $i=1,2,3$. The link between satellites is denoted as ``Link $i$",  and ``$L_{i}$'' represents its length. The two photon detectors on SC$_{i}$ are simply denoted as $i$ and $i'$. The deformation induced by the GW on Link $k$ detected by photon detector $j$ is denoted as $H_{ij}(t)$,
\bea
\balign
H_{ij}(t)=\,&h^{\text{SSB}}_{+} (t)\times \xi_{+}(\hat{\mathbf{u}},\hat{\mathbf{v}},\hat{\mathbf{n}}_{ij})\\
&+h_{\times}^{\text{SSB}}(t) \times \xi_{\times}(\hat{\mathbf{u}},\hat{\mathbf{v}},\hat{\mathbf{n}}_{ij}),
\ealign
\label{strainssb}
\eea
where $\xi_{+}$ and $\xi_{\times}$ are the antenna pattern functions which can be expressed as
\bea
\balign
\xi_{+}(\hat{\mathbf{u}},\hat{\mathbf{v}},\hat{\mathbf{n}}_{ij})&=(\hat{\mathbf{u}}\cdot\hat{\mathbf{n}}_{ij})^{2} - (\hat{\mathbf{v}}\cdot\hat{\mathbf{n}}_{ij})^{2},\,\\
\xi_{\times}(\hat{\mathbf{u}},\hat{\mathbf{v}},\hat{\mathbf{n}}_{ij})&=
2\,(\hat{\mathbf{u}}\cdot\hat{\mathbf{n}}_{ij})\,(\hat{\mathbf{v}}\cdot\hat{\mathbf{n}}_{ij}).
\ealign
\eea

The single-link observables, which characterizing the laser frequency shift of the laser from spacecraft $s$ to spacecraft $r$ along Link $l$, can be written as $y_{sr}=(\nu_{r}-\nu_{s})/\nu$ (we omitted the $l$ in the subscript of $y$ related to link, because through the emitting and receiving spacecraft, we can determine the corresponding link). The expression can be further expressed as \cite{Vallisneri:2004bn,Krolak:2004xp,Marsat:2020rtl}
\begin{equation}
\balign
 y_{s r}=\frac{1}{2}\frac{1}{1-\hat{\mathbf{k}} \cdot \hat{\mathbf{n}}_{sr}}
\times\left[H_{sr}\left(t-L-\hat{\mathbf{k}} \cdot \mathbf{x}_s\right)-H_{sr}\left(t-\hat{\mathbf{k}} \cdot \mathbf{x}_r\right)\right]\text{,}\\
\ealign
\label{ysr}
\end{equation}
where the $\mathbf{x}_{s}$ and $\mathbf{x}_{r}$ is the positions of the spacecraft $s$ and the spacecraft $r$ respectively, $\hat{\mathbf{n}}_{sr}$ is the unit vector of link (from $s$ to $r$), $\hat{\mathbf{k}}$ is the propagation vector for GW and $L$ is the length of link between spacecraft $s$ and $r$.

In this study, we employ the first generation TDI, which adopts the assumption of equal arm length, meaning that the delay on each link is constant, i.e., $L$. The first generation TDI Michelson combinations are given by \cite{Tinto:2003vj,Vallisneri:2004bn}
\begin{equation}
\balign
X&=  y_{31'}+\mathbf{D}_{3'1}\, y_{1'3}+\mathbf{D}_{31'}\, \mathbf{D}_{1'3} \,y_{2'1}+\mathbf{D}_{31'}\, \mathbf{D}_{1'3}\, \mathbf{D}_{2'1}\, y_{12'} \\
& -\left[y_{2'1}+\mathbf{D}_{2'1}\, y_{12'}+\mathbf{D}_{12'}\, \mathbf{D}_{2'1} \,y_{31'}+\mathbf{D}_{2'1} \,\mathbf{D}_{12'} \,\mathbf{D}_{31'}\, y_{1'3}\right].\\
\ealign
\label{tdiX}
\end{equation}
The operator $\mathbf{D}_{ij}$ is the delay operator which is defined by
\bea
\mathbf{D}_{ij}\,x(t) = x\,(t-L),
\eea
The other Michelson combinations, i.e., $Y$ and $Z$ can be obtained by cyclic permutation of indices: $1 \to 2 \to 3 \to 1$ in Eq. (\ref{tdiX}). The TDI combinations A, E and T can be constructed by these Michelson combinations as
\bea
\balign
& A=\frac{1}{\sqrt{2}}(Z-X), \\
& E=\frac{1}{\sqrt{6}}(X-2 Y+Z), \\
& T=\frac{1}{\sqrt{3}}(X+Y+Z) .
\ealign
\eea
We note that A, E, and T are orthogonal only in a situation of equal length.

\subsection{Parameter Estimation method} 
According to Bayes theorem, the posterior probability distribution $p(\boldsymbol{\theta}\,|\, d, D)$ of GW source parameters $\boldsymbol{\theta}$ can be expressed as
\bea
p(\boldsymbol{\theta} \,|\, d, \mathcal{M})=\frac{p(d \,|\, \boldsymbol{\theta}, \mathcal{M})\, p(\boldsymbol{\theta}\,|\, \mathcal{M})}{p(d \,|\,\mathcal{M})},
\label{bayestherom}
\eea
where $d$ is data, which is GW signal for the true parameters $\boldsymbol{\theta}_0$ with noise realization in the experiment $n$ as measured on the detector, i.e., $d=h(\boldsymbol{\theta}_0)+n$. $\mathcal{M}$ is the assumed model for the signal, and the source parameters $\boldsymbol{\theta}$ are associated with it. $p(d \,|\, \boldsymbol{\theta}, \mathcal{M}) = \mathcal{L}$ is the likelihood, representing the probability of the data given by the chosen model and the model parameters. $p(\boldsymbol{\theta}\,|\, \mathcal{M})$ is the prior probability on the parameters, and $p(d \,|\, D)=\mathcal{Z}$ denotes the evidence or marginalized likelihood. For stationary Gaussian noise, the likelihood function $\mathcal{L}$ of GW signals can be expressed as
\bea
\begin{aligned}
\log \mathcal{L} \propto & -\frac{1}{2}\langle d-h \mid d-h\rangle \\
& =-\frac{1}{2}(\langle d \mid d\rangle+\langle h \mid h\rangle-2\langle d \mid h\rangle).
\end{aligned}
\label{likelihood}
\eea
The $\langle ...\,|\, ...\rangle$ represents the inner product. The inner product of $a$ and $b$ can be written as
\bea
\langle a \mid b\rangle=4 \operatorname{Re} \int_0^{\infty} \frac{\tilde{a}^{*}(f) \tilde{b}(f)}{S_n(f)} \mathrm{d} f,
\label{inner}
\eea
where $\tilde{a}(f)$ is the fourier transform of $a(t)$ and $\tilde{a}^{*}(f)$ is the complex conjugate of $\tilde{a}(f)$. $S_{n}(f)$ is the one-sided power spectral density (PSD) of the detector noise. 

The primary goal of this study is to investigate the impact of neglecting memory on parameter estimation results and to explore the threshold for SNR in detecting memory by TianQin. However, random noise processes can induce shifts in the likelihood surface. To calculate the likelihood more accurately, facilitating an efficient exploration of the structure of the likelihood and the degeneracies between parameters, we adopt the zero-noise approximation which means set $n=0$ in the data \cite{Marsat:2020rtl,Katz:2022yqe,Toubiana:2020cqv,Lyu:2023ctt}.

 We adopt the one-sided PSD of TianQin for TDI channels, i.e., A, E, and T, which can be defined as
\bea
\begin{aligned}
S_n^A= & S_n^E=8 \sin ^2 2 \pi f L\left[4\left(1+\cos 2 \pi f L+\cos ^2 2 \pi f L\right) S_{\mathrm{acc}}\right. \\
& \left.+(2+\cos 2 \pi f L) S_{\mathrm{pos}}\right], \\
S_n^T= & 32 \sin ^2 2 \pi f L \sin ^2 \frac{2 \pi f L}{2}\left[4 \sin ^2 \frac{2 \pi f L}{2} S_{\mathrm{acc}}+S_{\mathrm{pos}}\right],
\end{aligned}
\eea
where the noise parameters $S_{\text{acc}}$ and $S_{\text{pos}}$ of TianQin can be found in Ref. \cite{TianQin:2015yph}.

The optimal SNR $\rho$ of GW signal $\tilde{s}(f)$ generated on the GW detector is defined as the square root of the inner product,
\bea
\rho = \sqrt{\langle \tilde{s}\,|\, \tilde{s}\rangle},\label{snr}
\eea
where $\tilde{s}$ is the Fourier transform of any among the channels $A$, $E$, and $T$.
 
In addition to calculating the posterior probability distribution of the parameters, we can also investigate the issue of model selection, namely, studying which model is more likely to preferred by the observed data. This is typically achieved by computing the Bayes factor between the models $\mathcal{M}_1$ and $\mathcal{M}_2$. The Bayes factor $\text{BF}^{1}_{2}$ for $\mathcal{M}_1$ and $\mathcal{M}_2$ is defined as
\bea
\text{BF}^{1}_{2}=\frac{\mathcal{Z}_{1}}{\mathcal{Z}_{2}},
\eea
where $\mathcal{Z}_{i}$ represents the evidence for the model $\mathcal{M}_{i}$ and can be calculated as
\bea
\mathcal{Z}_{i}=\int \mathcal{L}(d\,|\,\mathbf{\theta},\mathcal{M}_{i}) \,p(\theta\,|\,\mathcal{M}_{i}) \,d \mathbf{\theta}.
\eea
If the computed Bayes factor is positive, we consider the data to favor $\mathcal{M}_1$, and similarly, if it is negative, the data are more favorable to $\mathcal{M}_2$. The $\text{log}_{10}$ Bayes factor is commonly used in practice and can be written as 
\bea
\text{log}_{10}\, \text{BF}^{1}_{2} = \text{log}_{10}\,\mathcal{Z}_{1} - \text{log}_{10}\,\mathcal{Z}_{2}.
\label{logbayes}
\eea
We use a threshold of $\text{log}_{10}\, \text{BF}=8$ \cite{Lasky:2016knh,Thrane:2018qnx}, meaning that when $\text{log}_{10}\, \text{BF}=8$, the data strongly supports either model 1 or 2 (depending on the sign of the value). Furthermore, some studies use $\text{log}_{10}\, \text{BF}=3$ as strong evidence and $\text{log}_{10} \,\text{BF}=5$ as a detection threshold \cite{Goncharov:2023woe}. In our scenario, when the value of the $\text{log}_{10}$ Bayes factor ($\text{log}_{10}\, \text{BF}^{\,\text{mem}}_{\,\text{no mem}}$) between conditions with and without memory reaches 8, we consider that the data contain memory, indicating the detectability of memory.

\section{Code implementation and data simulation}\label{sec4}
This work is a part of the TianQin data analysis pipeline, intended to complement the existing parameter estimation program implemented with emcee and the frequency TDI response of TianQin \cite{Lyu:2023ctt}. We aim to compute memory waveforms incorporating the time-domain TDI response of TianQin and to perform parameter estimation and Bayesian evidence calculation for signals with memory generated by the merger of MBHBs observed by TianQin. 

For Bayesian inference algorithm, we use dynesty \cite{Speagle:2019ivv,sergey_koposov_2023_8408702}, which is based on the nested sampling method \cite{10.1063/1.1835238,Skilling:2006gxv}. Because dynesty excels in exploring the shape of the likelihood, particularly adept at effectively capturing the multi-modal structure of posterior probability distributions. Additionally, dynesty can simultaneously calculate Bayesian evidence and the posterior probability distributions of parameters.

We use the IMRPhenomXHM waveform model \cite{Pratten:2020fqn,Garcia-Quiros:2020qpx,Colleoni:2020tgc}, which is a model with aligned spins and can rapidly generate relatively accurate frequency-domain GW waveforms. Additionally, it provides several dominant higher-order modes of gravitational waves, namely (2,2), (2,1), (3,3), (3,2), and (4,4) modes. 

The waveform of the memory effect currently needs to be computed in the time domain, and we obtain the time-domain IMRPhenomXHM waveform through the inverse Fourier transform method built into LALsuite \cite{LALSuite}. Subsequently, we convert the time-domain waveforms to the frequency domain by applying a Plank-taper window \cite{McKechan:2010kp} to calculate the likelihood by using Eq. (\ref{likelihood}).

Previous studies have indicated that the SNR of GW signals produced by the merger of MBHB on TianQin primarily originates approximately one day before the merger \cite{haitian}. To optimize computational efficiency and ensure the validity of the results, we choose a simulation signal length of two days. Regarding total mass and mass ratio, another commonly used representation is the chirp mass $M_{c}$ and the symmetric mass ratio $\eta$. The chirp mass $M_{c}$ is defined as
\bea
M_{c}=\frac{(m_{1}m_{2})^{3/5}}{(m_1 + m_2)^{1/5}},
\eea
where $m_1$ and $m_2$ are the masses of two components of binaries. The symmetric mass ratio $\eta$ is defined as
\bea
\eta = \frac{m_1 m_2}{(m_1 + m_2)^2}.
\eea
\begin{table}[t]
\setlength{\tabcolsep}{3mm}{
\begin{tabular}{lcc}
\hline \hline
Parameter &  Symbol  & Value \\
\hline
Source-frame chirp mass ($\text{M}_{\odot}$) & $M_{c}$ & 385373 \\
\hline
Symmetric mass ratio & $\eta$ &  0.204   \\
\hline
Spin 1, Spin 2 & ($\chi_1,\,\chi_2$) & (0.4,\,0.2)  \\
\hline
Luminosity distance (Gpc) & $D_L$ & 10  \\
\hline
Redshift & $z$ & 1.37\\
\hline
Inclination (rad) & $\iota$ & $\pi/3$   \\
\hline
Reference phase (rad) & $\varphi$ & 0  \\
\hline
Ecliptic longitude (rad)  & $\lambda$ & $\pi/4$ \\
\hline
Ecliptic latitude (rad)  & $\beta$ & $\pi/5$ \\
\hline
Polarization angle (rad)  & $\psi$ & $\pi/2$ \\
\hline
\hline
\end{tabular}}
\caption{Parameters of simulated MBHB system.}
\label{parasum}
\end{table}

\begin{figure}[t]
\subfigure{
    \centering
    \includegraphics[scale=0.5]{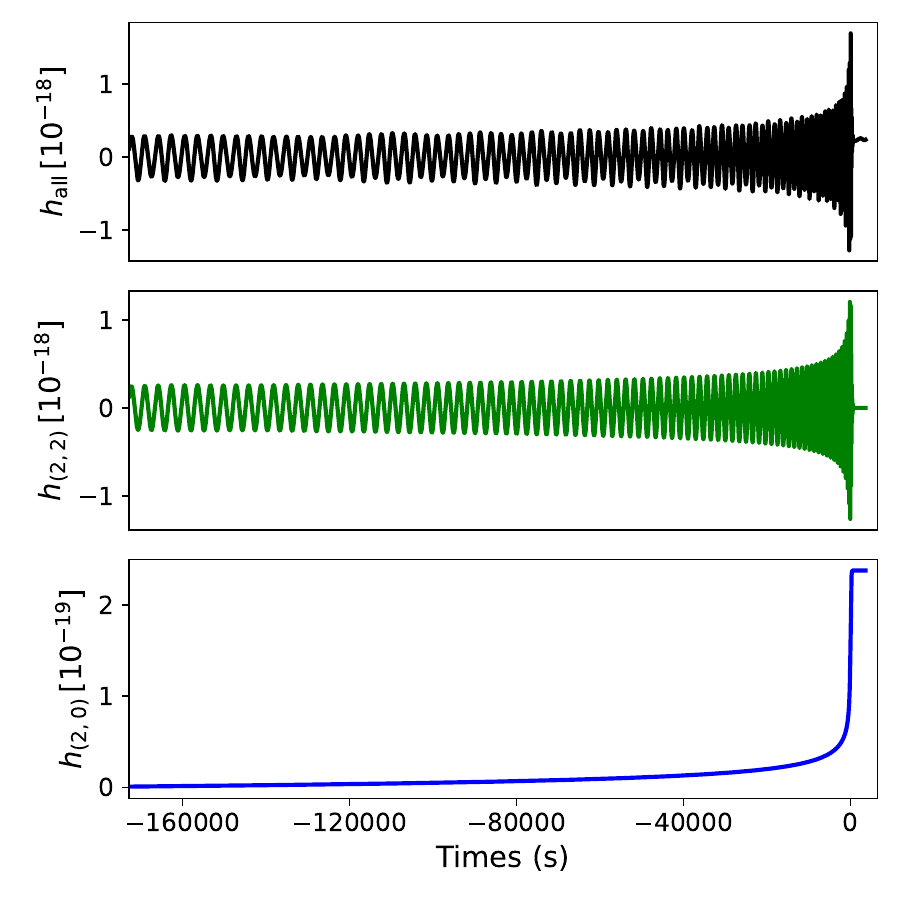}
    }
\subfigure{
    \centering
    \includegraphics[scale=0.45]{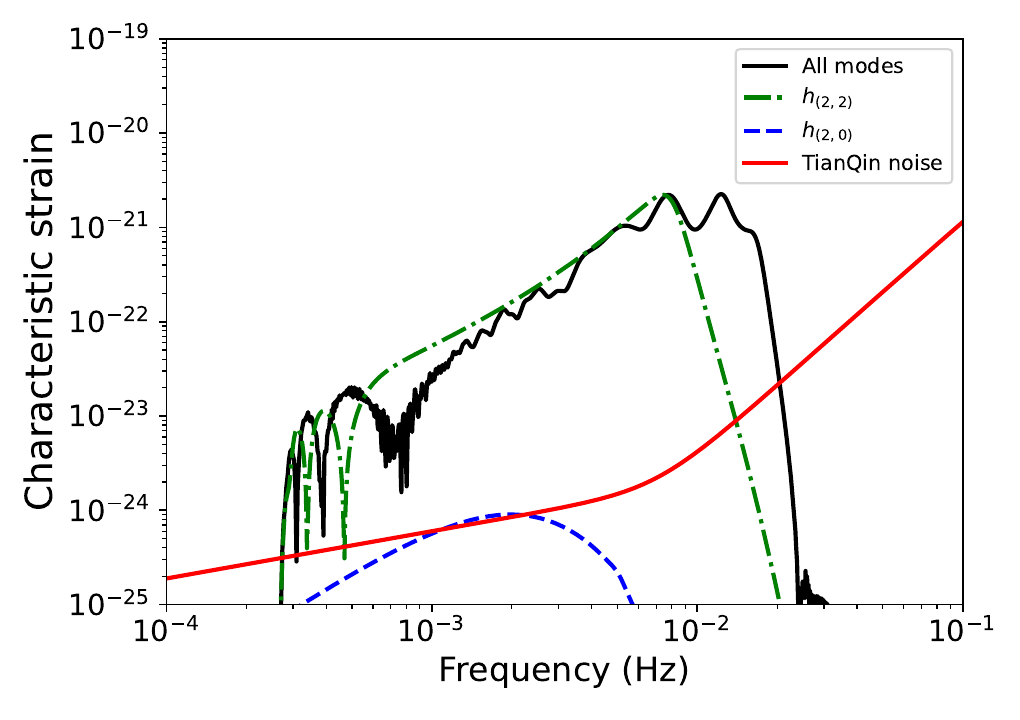}
    }
    \caption{The GW strains for the simulated MBHB system with the parameters listed in the Table. \ref{parasum}. The top panel shows the time-domain strains and the lower panel shows the characteristic strains $2f\,|\tilde{h}(f)|$ for TDI A channel in the frequency domain together with TianQin Noise $\sqrt{f\,S_{n}(f)}$ (red solid line).  Three main strains are considered: $h_{\text{all}} $ (black), (2,2) mode $h_{(2,2)}$ (green), and the dominant memory mode $h_{(2,0)}$ (blue).}
    \label{strains}
\end{figure}

 All the parameters of the simulated MBHB system are listed in the Table. \ref{parasum}. By this code, we generate the A and E channels of the simulated data, and we neglect the T channel due to its low-frequency insensitivity. In the upper panel of Fig. \ref{strains}, we plot the time-domain strains for the sum of all modes of GW signals $h_{\text{all}}$ (black), the dominant mode $h_{(2,2)}$ (green), and the dominant memory mode $h_{(2,0)}$ (blue). The lower panel shows the characteristic strain $\tilde{h}_{c}(f)=2f\,|\tilde{h}(f)|$ of the A channel for these modes, and $\tilde{h}(f)$ represents the Fourier transform of time-domain strain $h(t)$. We only plot the A channel because channel E is similar to that of channel A. The solid red line in Fig. \ref{strains} represents the TianQin noise $\sqrt{f\, S_{n}(f)}$. The SNR for the total modes of the GW signal is $\rho_{\text{all}}\approx 1058.03$ and the memory SNR is $\rho_{\text{mem}}\approx 1.64$. 


\section{Results}\label{sec5}
\begin{table*}[t]
\setlength{\tabcolsep}{2mm}{
\begin{tabular}{lcccc}
\hline \hline
Parameter &  Symbol  & Prior & With memory & Without memory\\
\hline
Source-frame chirp mass ($\text{M}_{\odot}$) & $M_{c}$ & [387373, 383373] & $385379.70009^{+86.0264}_{-85.1294}$ & $385389.6932^{+92.1852}_{-80.3023}$ \\
\hline
Symmetric mass ratio & $\eta$ &  [0.15, 0.25] &  $0.2040^{+0.0002}_{-0.0001}$ & $0.2040^{+0.0001}_{-0.0002}$  \\
\hline
Spin 1 & $\chi_a$ & [0,\,0.5] & $0.3003^{+0.0018}_{-0.0020}$ & $0.3004^{+0.0021}_{-0.0021}$  \\
\hline
Spin 2 & $\chi_l$ & [0,\,0.5] & $0.0996^{+0.0025}_{-0.0024}$ & $0.0996^{+0.0026}_{-0.0028}$  \\
\hline
Luminosity distance (Mpc) & $D_{\text{L}}$ & [9000, 11000] & $10013.8591^{+54.7595}_{-55.6548}$ & $10060.2307^{+51.2933}_{-54.6047}$ \\
\hline
Inclination (rad) & $\iota$ & $[0,\, \pi]$  & $1.0465^{+0.0059}_{-0.0059}$ & $2.0933^{+0.0090}_{-1.0515}$ \\
\hline
Reference phase (rad) & $\varphi$ & $[0, \,2\pi]$ & $0.0024^{+0.0186}_{-0.0183}$ & $0.0049^{+0.0218}_{-0.0187}$\\
\hline
Ecliptic longitude (rad)  & $\lambda$ & $[0,\,2\pi]$ & $0.6283^{+0.0008}_{-0.0007}$ & $0.6283^{+0.0008}_{-0.0007}$ \\
\hline
Ecliptic latitude (rad)  & $\beta$ & $[-\frac{\pi}{2},\frac{\pi}{2}]$ & $0.7852^{+0.0009}_{-0.0007}$ & $0.7849^{+0.0009}_{-0.0007}$\\
\hline
Polarization angle (rad)  & $\psi$ & $[0,\,\pi]$ & $1.5706^{+0.0054}_{-0.0057}$ & $1.5711^{+0.0066}_{-0.0055}$ \\
\hline
\hline
\end{tabular}}
\caption{Priors and parameter estimation results for the simulated MBHB's parameters. The last two columns show the 1$\sigma$ confidence region of parameters for the GW signal with and without memory from the simulated MBHB merger.}
\label{prior}
\end{table*}
\subsection{Impact of memory on parameter estimation}
We consider the ideal scenario in which no other signals are mixed; that is, only the GW signal from the simulated MBHB merger is injected into the data.

To Obtain posterior distributions of parameters, one needs to determine the prior probability of all parameters. We adopt uninformative priors, which means that the prior distributions of all parameters are flat. In our scenario, the presence of a strong correlation between the two spins $(\chi_1, \chi_2)$ leads to a decrease in the sampling efficiency and introduces errors in the estimation of other parameters. Therefore, we reparametrize the two spins as
\bea
\balign
\chi_a = \frac{\chi_1 + \chi_2}{2},\\
\chi_l=\frac{\chi_1 - \chi_2}{2}.
\ealign
\eea
Throughout the entire parameter estimation process, for the spin, we choose to sample $\chi_a$ and $\chi_l$.

We assume that the detection pipeline has successfully identified the signal, therefore, we set the priors of intrinsic parameters and the luminosity distance of the signal, i.e., $\{M_c,\,\eta,\,\chi_a,\,\chi_l,\, D_{\text{L}}\}$, to be around their injected values. For extrinsic parameters, i.e., $\{\iota,\,\varphi,\,\lambda,\,\beta,\psi\}$, we sample over the entire parameter space. All priors for parameters of simulated MBHB are listed in the third column of Table. \ref{prior}.

\begin{figure*}[t]
    \centering
    \includegraphics[scale=0.32]{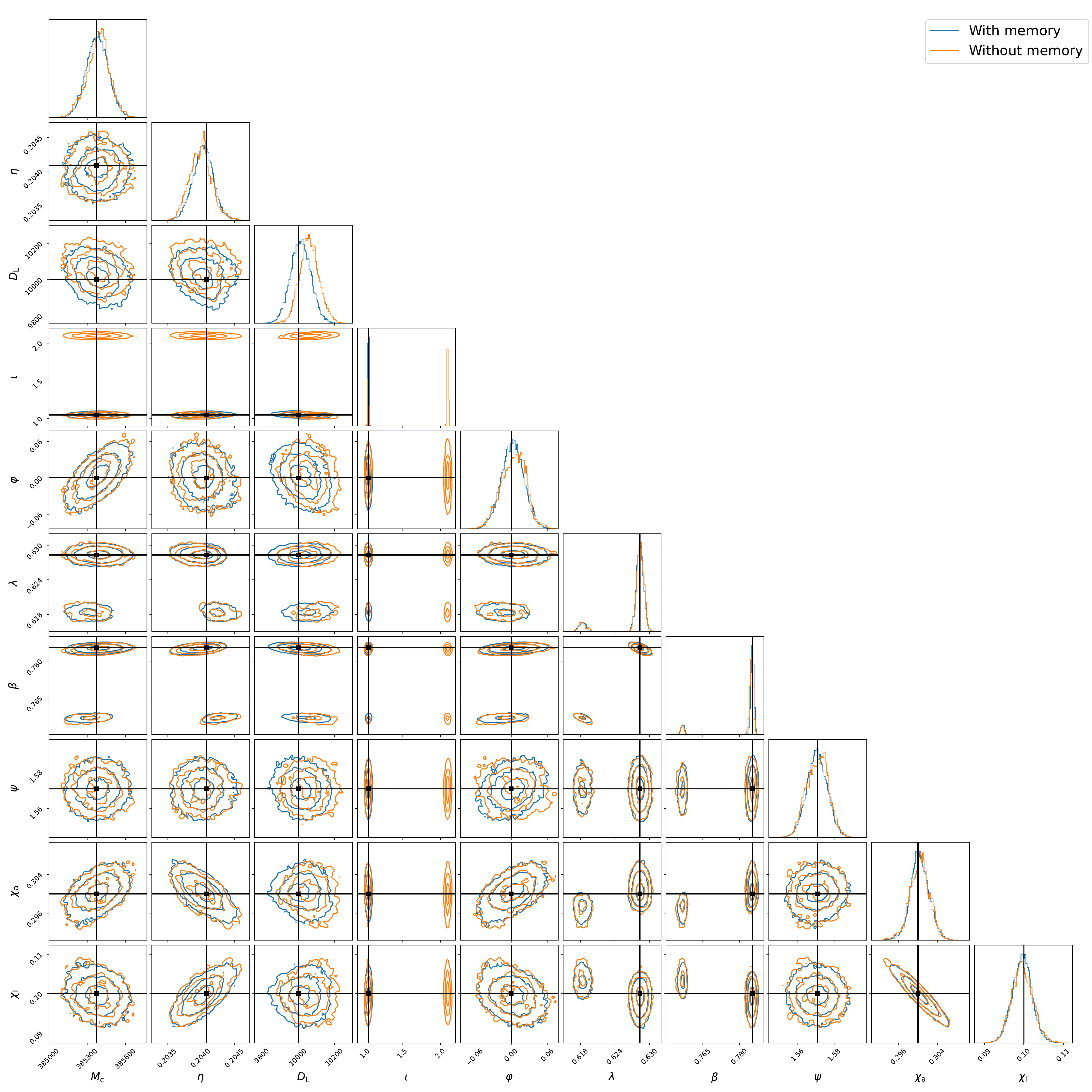}
    \caption{The parameter estimation results of GW signals with and without memory for TianQin. The blue curve represents signals with memory, while the orange curve represents signals without memory. The solid black line indicates the true values of the parameters, and the contour plots represent the $1-3\sigma$ confidence regions of the inferred parameter estimation results.}
    \label{corner1}
\end{figure*}

We perform parameter estimation on simulated signal using waveforms with memory (denoted as ``with memory'') and waveforms without memory (denoted as ``without memory"). The constraint results of both ``with memory'' and ``without memory'' along with their $1\sigma$ confidence regions for parameters of the simulated signal are presented in 
 the last two columns of Table. \ref{prior}. The corner plot that represents the overlaid posterior distribution of ``with memory'' and ``without memory'' for the parameters is shown in Fig. \ref{corner1}. One can see from the last two columns of the Table \ref{prior}, the parameter estimation precision for $M_{c}$ and $D_{\text{L}}$ obtained through dynesty can reach approximately $10^2$, while for other parameters, the precision is typically around $10^{-2}$ to $10^{-4}$.

Due to the injected value $\pi/3$ for $\iota$, it induces a symmetric structure about $\pi/2$ on the likelihood surface. Consequently, the posterior for $\iota$ exhibits a bimodal distribution. During the computational process of the program, the sampling may ``jump'' between two modes, eventually causing some modes to ``die off'', i.e., the samples are completely focused on one mode, leading to the disappearance of the other mode. This issue will make it challenging to achieve a completely identical distribution in the final results. In our results, we obtain samples from the bimodal distribution in the case of ``without memory''. This is why in Table \ref{prior}, for the inclination in the case ``without memory'', there is a significant fluctuation in the $1\sigma$ confidence region. In the case of ``with memory'', the posterior distribution of the inclination exhibits only one mode, corresponding to the injected value $\pi/3$. We believe this is attributed to mode ``die-off'', causing the mode at $2\pi/3$ not to be sampled. The multi-modal distributions of posteriors for $\lambda$ and $\beta$ are also illustrated in Fig. \ref{corner1}. Regarding $\beta$ and $\lambda$, two additional modes emerge, with the number of samples that fall into them during the sampling process being so small that only two contours are visible in the figure.

Because memory is a low-frequency effect with relatively weak intensity, its impact on the GW signal is highly limited. Our full Bayesian analysis of the signal with memory suggests that when using waveforms without memory for parameter estimation, most parameters do not exhibit significant deviations.

For parameters $D_{\text{L}}$ and $\iota$, using waveforms without memory results in a significant deviation from their parameter estimation result, with the injected values deviating from the $1\sigma$ confidence region. Therefore, in the case of high SNR for GW signals from MBHBs, neglecting memory may significantly affect the parameter estimation of $D_{\text{L}}$. Some studies suggest that introducing memory can break the degeneracy between $D_{\text{L}}$ and $\iota$ \cite{Gasparotto:2023fcg}. However, for TianQin, since it is less sensitive to low-frequency signals compared to LISA, our results do not exhibit a significant break in the degeneracy.

\subsection{The detection threshold and waveform mismodeling}
Despite the limited impact of memory on the parameter estimation results of GW signals generated by the merger of MBHB for TianQin, detecting memory remains a worthwhile and intriguing pursuit. We investigated the SNR threshold to detect memory effects with TianQin through the Bayesian factor.

 For the detected GW data, we evaluate the Bayesian evidence for both the ``with memory'' hypothesis and the ``without memory'' hypothesis using waveforms with and without memory respectively and calculate the $\text{log} _{10}$ Bayes factor by using Eq. (\ref{logbayes}). As mentioned earlier, we set the threshold for the $\text{log}_{10} \,\text{BF}^{\,\text{mem}}_{\,\text{no mem}}$ to 8 \cite{Lasky:2016knh}. In other words, when the $\text{log}_{10} \,\text{BF}^{\text{\,mem}}_{\text{\,no mem}}$ reaches 8, we can confidently say that the memory has been detected.

\begin{figure}
    \centering
    \includegraphics[scale=0.45]{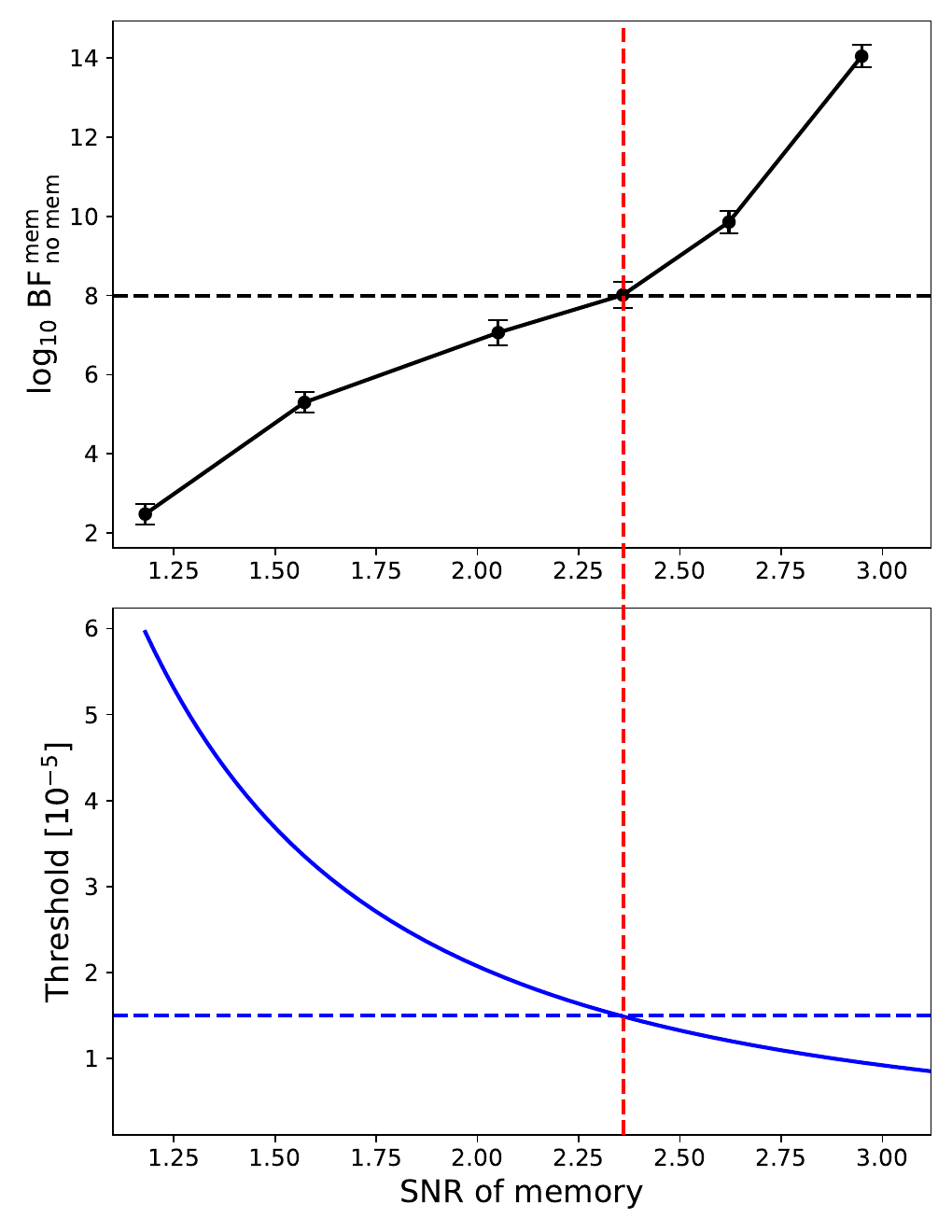}
    \caption{The $\text{log}_{10}$ Bayes factor $\text{log}_{10} \text{BF}^{\text{mem}}_{\text{no mem}}$ and the threshold of mismatch as functions of memory SNR on TianQin. The horizontal black dashed line represents the $\text{log}_{10} \text{BF}^{\text{mem}}_{\text{no mem}}$ equal to 8 and the horizontal blue dashed line represents the mismatch between waveforms with and without memory. The red vertical dashed line represents the SNR of approximately 2.36 for the memory.}
    \label{threshold}
\end{figure}

The results of $\text{log}_{10}\text{BF}^{\,\text{mem}}_{\,\text{with mem}}$ as a function of memory SNR are shown in upper panel of Fig. \ref{threshold}, these results were obtained by performing 12 independent samplings using dynesty for both the waveform with memory and the waveform without memory. The dashed red line indicates the threshold of the $\text{log}_{10}$ Bayes factor equal to 8. At this threshold, the corresponding SNR for memory is approximately 2.36. This implies that if the SNR for memory generated by TianQin reaches around 2.36, TianQin would be able to detect the memory.

Another topic is waveform mismodeling. Currently, the most commonly used GW waveform models do not include memory. Therefore, using waveforms without memory for parameter estimation will introduce systematic error due to waveform mismodeling. In other words, it can be argued that if the systematic errors introduced by neglecting memory in the waveform model become significant, one can intuitively reason that this should be equivalent to when the memory will become detectable.

The effects arising from neglecting memory in waveform models can be quantified through the mismatch between the waveform models with and without memory. For two different waveform models $\tilde{h}_{1}(f)$ and $\tilde{h}_{2}(f)$, the mismatch is defined as 
\bea
\mathcal{M}=1-\frac{\left\langle\tilde{h}_1, \tilde{h}_2\right\rangle}{\sqrt{\left\langle\tilde{h}_1, \tilde{h}_1\right\rangle\left\langle\tilde{h}_2, \tilde{h}_2\right\rangle}},
\eea
where $\left\langle\tilde{h}_1, \tilde{h}_2\right\rangle$ is the inner product for a given detector PSD of two waveforms and defined in Eq. (\ref{inner}).

In high SNR regions, statistical errors can be estimated by the inverse of Fisher matrix $\Gamma^{ij}(\partial_i h|\partial_j h)$ (where $i$, $j$ are the waveform parameters), and these errors are decreased as SNR$^{-1}$ \cite{Vallisneri:2007ev}. The threshold for the mismatch is used to assess whether the systematic errors introduced by incorrect waveform models would impact parameter estimation, which is defined as \cite{Baird:2012cu,Chatziioannou:2017tdw,Mangiagli:2018kpu,Purrer:2019jcp,Toubiana:2023cwr}
\bea
\mathcal{M}_{\text{threshold}}=\frac{D}{2\, \text{SNR}^2}.
\label{misthre}
\eea
For a given SNR and PSD, the systematic errors from waveform inaccuracies are expected to be smaller than the statistical errors when mismatch $\mathcal{M}$ smaller than this threshold. The factor $D$ is not known accurately, but it is generally estimated as the number of intrinsic parameters whose estimation is affected by the waveform accuracy \cite{Chatziioannou:2017tdw}, or it can be tuned by calculating the statistical and systematic errors from the posterior distribution of synthetic signals with increasing SNRs \cite{Purrer:2019jcp}. Since IMRPhenomXHM is an aligned spin waveform model, in our study, we take $D=4$. Some studies indicate that the threshold calculated by Eq. (\ref{misthre}) is generally too conservative, and when it is violated, biases do not necessarily appear in parameter estimation results \cite{Pompili:2023tna,Ossokine:2020kjp}.

In the lower panel of Fig. \ref{threshold}, we show that the mismatch threshold $\mathcal{M}_{\text{threshold}}$ varies with the increase in the SNR of memory on TianQin.  From our results, we find that when the SNR of memory reaches approximately 2.36, the mismatch becomes equal to the threshold value. As the SNR of memory continues to increase, the threshold will be less than the threshold and introduce biases in parameter estimation due to the use of waveforms without memory. Furthermore, the SNR of memory when the mismatch equals the threshold is consistent with the SNR corresponding to $\text{Log BF}^{\,\text{mem}}_{\,\text{no mem}}=8$. This implies that when TianQin can detect memory, it may lead to biases in parameter estimation due to the use of waveforms without memory. However, due to the relatively low SNR of memory, even when the mismatch exceeds the threshold, no significant bias is observed in many parameters. Only the result of the parameter estimation for $D_L$ and $\iota$ deviate beyond the $1\sigma$ confidence region. This can be seen in Fig. \ref{corner1}.
\section{Conclusion}\label{conclusion}
General relativity predicts that gravitational radiation can permanently change the background spacetime. This change arises not only from variations in BMS charges but also from the energy and angular momentum flux radiated to null infinity. This distinctive effect is referred to as the gravitational wave memory effect. Consequently, the detection of the gravitational wave memory effect is crucial for advancing our understanding of the nature of spacetime and gravity.

In the future, space-based gravitational wave detection projects are expected to achieve higher precision, enabling the detection of GW signals produced by the merger of MBHBs. These signals are expected to be more intense and radiate more energy and angular momentum outward during the GW emission. This advancement makes it possible to directly detect memory effects in individual events. 

In this study, we carried out Bayesian inference on the memory in GW signal generated by simulated MBHB detected by TianQin. The code utilizes dynesty to compute posterior probability and is capable of applying TianQin TDI response in the time domain while generating memory waveforms. Through the calculations performed by this code, we observed that neglecting memory does not introduce significant biases in most signal parameters but does impact $D_{\text{L}}$ and $\iota$. Calculating the Bayes factors revealed that when the SNR of memory on TianQin reaches approximately 2.36, the $\text{log}_{10}$ Bayes factor reaches 8, indicating that memory is detectable. Additionally, our findings show consistency between the results obtained through $\text{log}_{10}$ Bayes factor and mismatch threshold. This suggests that the SNR of memory, given by the mismatch equal to the threshold, can serve as a detection threshold to some extent. However, due to a limited understanding of the mismatch threshold \cite{Purrer:2019jcp,Toubiana:2023cwr}, more extensive and detailed research is needed to further investigate this issue.

As an initial exploration of parameter estimation for memory, our study is built on many ideal assumptions, including no signal overlap and zero-noise assumptions. However, in real-world scenarios, when multiple signals overlap, memory may also superimpose. This raises questions about whether parameter estimation results are affected and whether memory effects can be regarded as a signal akin to a stochastic gravitational wave background. This is a very interesting direction for further investigation. Furthermore, the investigation of the dependency of memory parameter estimation on various parameters and whether different waveform models may impact memory parameter estimation is equally worth exploring. This contributes to a more comprehensive understanding of memory detection and its potential implications on parameter estimation. We will leave these aspects for the future work.

\section{Acknowledgement}
We are greatful to Sergey Koposov for his kind helps in using dynesty. We thank Yi-Ming Hu, Han Wang, En-Kun Li, Xiangyu Lyu and Chang-Qing Ye for many useful discussions and helps. We acknowledge the usage of the calculation utilities of dynesty \cite{sergey_koposov_2023_8408702}, LALsuite \cite{LALSuite}, NUMPY \cite{vanderWalt:2011bqk}, and SCIPY \cite{Virtanen:2019joe}, and plotting utilities of MATPLOTLIB \cite{Hunter:2007ouj} and corner \cite{corner}. This work is supported by the Guangdong Major Project of Basic and Applied Basic Research (Grant No. 2019B030302001), the Guangdong Basic and Applied Basic Research Foundation (Grant No. 2023A1515030116),  and the National Science Foundation of China (Grant No. 12261131504). 
\bibliographystyle{apsrev4-1}
\bibliography{membayes}

\begin{thebibliography}{116}%
\makeatletter
\providecommand \@ifxundefined [1]{%
 \@ifx{#1\undefined}
}%
\providecommand \@ifnum [1]{%
 \ifnum #1\expandafter \@firstoftwo
 \else \expandafter \@secondoftwo
 \fi
}%
\providecommand \@ifx [1]{%
 \ifx #1\expandafter \@firstoftwo
 \else \expandafter \@secondoftwo
 \fi
}%
\providecommand \natexlab [1]{#1}%
\providecommand \enquote  [1]{``#1''}%
\providecommand \bibnamefont  [1]{#1}%
\providecommand \bibfnamefont [1]{#1}%
\providecommand \citenamefont [1]{#1}%
\providecommand \href@noop [0]{\@secondoftwo}%
\providecommand \href [0]{\begingroup \@sanitize@url \@href}%
\providecommand \@href[1]{\@@startlink{#1}\@@href}%
\providecommand \@@href[1]{\endgroup#1\@@endlink}%
\providecommand \@sanitize@url [0]{\catcode `\\12\catcode `\$12\catcode `\&12\catcode `\#12\catcode `\^12\catcode `\_12\catcode `\%12\relax}%
\providecommand \@@startlink[1]{}%
\providecommand \@@endlink[0]{}%
\providecommand \url  [0]{\begingroup\@sanitize@url \@url }%
\providecommand \@url [1]{\endgroup\@href {#1}{\urlprefix }}%
\providecommand \urlprefix  [0]{URL }%
\providecommand \Eprint [0]{\href }%
\providecommand \doibase [0]{http://dx.doi.org/}%
\providecommand \selectlanguage [0]{\@gobble}%
\providecommand \bibinfo  [0]{\@secondoftwo}%
\providecommand \bibfield  [0]{\@secondoftwo}%
\providecommand \translation [1]{[#1]}%
\providecommand \BibitemOpen [0]{}%
\providecommand \bibitemStop [0]{}%
\providecommand \bibitemNoStop [0]{.\EOS\space}%
\providecommand \EOS [0]{\spacefactor3000\relax}%
\providecommand \BibitemShut  [1]{\csname bibitem#1\endcsname}%
\let\auto@bib@innerbib\@empty
\bibitem [{\citenamefont {Abbott}\ \emph {et~al.}(2019{\natexlab{a}})\citenamefont {Abbott} \emph {et~al.}}]{LIGOScientific:2018mvr}%
  \BibitemOpen
  \bibfield  {author} {\bibinfo {author} {\bibfnamefont {B.~P.}\ \bibnamefont {Abbott}} \emph {et~al.} (\bibinfo {collaboration} {LIGO Scientific, Virgo}),\ }\href {\doibase 10.1103/PhysRevX.9.031040} {\bibfield  {journal} {\bibinfo  {journal} {Phys. Rev. X}\ }\textbf {\bibinfo {volume} {9}},\ \bibinfo {pages} {031040} (\bibinfo {year} {2019}{\natexlab{a}})},\ \Eprint {http://arxiv.org/abs/1811.12907} {arXiv:1811.12907 [astro-ph.HE]} \BibitemShut {NoStop}%
\bibitem [{\citenamefont {Abbott}\ \emph {et~al.}(2021{\natexlab{a}})\citenamefont {Abbott} \emph {et~al.}}]{LIGOScientific:2020ibl}%
  \BibitemOpen
  \bibfield  {author} {\bibinfo {author} {\bibfnamefont {R.}~\bibnamefont {Abbott}} \emph {et~al.} (\bibinfo {collaboration} {LIGO Scientific, Virgo}),\ }\href {\doibase 10.1103/PhysRevX.11.021053} {\bibfield  {journal} {\bibinfo  {journal} {Phys. Rev. X}\ }\textbf {\bibinfo {volume} {11}},\ \bibinfo {pages} {021053} (\bibinfo {year} {2021}{\natexlab{a}})},\ \Eprint {http://arxiv.org/abs/2010.14527} {arXiv:2010.14527 [gr-qc]} \BibitemShut {NoStop}%
\bibitem [{\citenamefont {Abbott}\ \emph {et~al.}(2021{\natexlab{b}})\citenamefont {Abbott} \emph {et~al.}}]{LIGOScientific:2021djp}%
  \BibitemOpen
  \bibfield  {author} {\bibinfo {author} {\bibfnamefont {R.}~\bibnamefont {Abbott}} \emph {et~al.} (\bibinfo {collaboration} {LIGO Scientific, VIRGO, KAGRA}),\ }\href@noop {} {\  (\bibinfo {year} {2021}{\natexlab{b}})},\ \Eprint {http://arxiv.org/abs/2111.03606} {arXiv:2111.03606 [gr-qc]} \BibitemShut {NoStop}%
\bibitem [{\citenamefont {Abbott}\ \emph {et~al.}(2019{\natexlab{b}})\citenamefont {Abbott} \emph {et~al.}}]{LIGOScientific:2019fpa}%
  \BibitemOpen
  \bibfield  {author} {\bibinfo {author} {\bibfnamefont {B.~P.}\ \bibnamefont {Abbott}} \emph {et~al.} (\bibinfo {collaboration} {LIGO Scientific, Virgo}),\ }\href {\doibase 10.1103/PhysRevD.100.104036} {\bibfield  {journal} {\bibinfo  {journal} {Phys. Rev. D}\ }\textbf {\bibinfo {volume} {100}},\ \bibinfo {pages} {104036} (\bibinfo {year} {2019}{\natexlab{b}})},\ \Eprint {http://arxiv.org/abs/1903.04467} {arXiv:1903.04467 [gr-qc]} \BibitemShut {NoStop}%
\bibitem [{\citenamefont {Abbott}\ \emph {et~al.}(2021{\natexlab{c}})\citenamefont {Abbott} \emph {et~al.}}]{LIGOScientific:2020tif}%
  \BibitemOpen
  \bibfield  {author} {\bibinfo {author} {\bibfnamefont {R.}~\bibnamefont {Abbott}} \emph {et~al.} (\bibinfo {collaboration} {LIGO Scientific, Virgo}),\ }\href {\doibase 10.1103/PhysRevD.103.122002} {\bibfield  {journal} {\bibinfo  {journal} {Phys. Rev. D}\ }\textbf {\bibinfo {volume} {103}},\ \bibinfo {pages} {122002} (\bibinfo {year} {2021}{\natexlab{c}})},\ \Eprint {http://arxiv.org/abs/2010.14529} {arXiv:2010.14529 [gr-qc]} \BibitemShut {NoStop}%
\bibitem [{\citenamefont {Abbott}\ \emph {et~al.}(2021{\natexlab{d}})\citenamefont {Abbott} \emph {et~al.}}]{LIGOScientific:2021sio}%
  \BibitemOpen
  \bibfield  {author} {\bibinfo {author} {\bibfnamefont {R.}~\bibnamefont {Abbott}} \emph {et~al.} (\bibinfo {collaboration} {LIGO Scientific, VIRGO, KAGRA}),\ }\href@noop {} {\  (\bibinfo {year} {2021}{\natexlab{d}})},\ \Eprint {http://arxiv.org/abs/2112.06861} {arXiv:2112.06861 [gr-qc]} \BibitemShut {NoStop}%
\bibitem [{\citenamefont {Abbott}\ \emph {et~al.}(2019{\natexlab{c}})\citenamefont {Abbott} \emph {et~al.}}]{LIGOScientific:2018jsj}%
  \BibitemOpen
  \bibfield  {author} {\bibinfo {author} {\bibfnamefont {B.~P.}\ \bibnamefont {Abbott}} \emph {et~al.} (\bibinfo {collaboration} {LIGO Scientific, Virgo}),\ }\href {\doibase 10.3847/2041-8213/ab3800} {\bibfield  {journal} {\bibinfo  {journal} {Astrophys. J. Lett.}\ }\textbf {\bibinfo {volume} {882}},\ \bibinfo {pages} {L24} (\bibinfo {year} {2019}{\natexlab{c}})},\ \Eprint {http://arxiv.org/abs/1811.12940} {arXiv:1811.12940 [astro-ph.HE]} \BibitemShut {NoStop}%
\bibitem [{\citenamefont {Abbott}\ \emph {et~al.}(2021{\natexlab{e}})\citenamefont {Abbott} \emph {et~al.}}]{LIGOScientific:2020kqk}%
  \BibitemOpen
  \bibfield  {author} {\bibinfo {author} {\bibfnamefont {R.}~\bibnamefont {Abbott}} \emph {et~al.} (\bibinfo {collaboration} {LIGO Scientific, Virgo}),\ }\href {\doibase 10.3847/2041-8213/abe949} {\bibfield  {journal} {\bibinfo  {journal} {Astrophys. J. Lett.}\ }\textbf {\bibinfo {volume} {913}},\ \bibinfo {pages} {L7} (\bibinfo {year} {2021}{\natexlab{e}})},\ \Eprint {http://arxiv.org/abs/2010.14533} {arXiv:2010.14533 [astro-ph.HE]} \BibitemShut {NoStop}%
\bibitem [{\citenamefont {Abbott}\ \emph {et~al.}(2021{\natexlab{f}})\citenamefont {Abbott} \emph {et~al.}}]{LIGOScientific:2021psn}%
  \BibitemOpen
  \bibfield  {author} {\bibinfo {author} {\bibfnamefont {R.}~\bibnamefont {Abbott}} \emph {et~al.} (\bibinfo {collaboration} {LIGO Scientific, VIRGO, KAGRA}),\ }\href@noop {} {\  (\bibinfo {year} {2021}{\natexlab{f}})},\ \Eprint {http://arxiv.org/abs/2111.03634} {arXiv:2111.03634 [astro-ph.HE]} \BibitemShut {NoStop}%
\bibitem [{\citenamefont {Isi}\ \emph {et~al.}(2019)\citenamefont {Isi}, \citenamefont {Giesler}, \citenamefont {Farr}, \citenamefont {Scheel},\ and\ \citenamefont {Teukolsky}}]{Isi:2019aib}%
  \BibitemOpen
  \bibfield  {author} {\bibinfo {author} {\bibfnamefont {M.}~\bibnamefont {Isi}}, \bibinfo {author} {\bibfnamefont {M.}~\bibnamefont {Giesler}}, \bibinfo {author} {\bibfnamefont {W.~M.}\ \bibnamefont {Farr}}, \bibinfo {author} {\bibfnamefont {M.~A.}\ \bibnamefont {Scheel}}, \ and\ \bibinfo {author} {\bibfnamefont {S.~A.}\ \bibnamefont {Teukolsky}},\ }\href {\doibase 10.1103/PhysRevLett.123.111102} {\bibfield  {journal} {\bibinfo  {journal} {Phys. Rev. Lett.}\ }\textbf {\bibinfo {volume} {123}},\ \bibinfo {pages} {111102} (\bibinfo {year} {2019})},\ \Eprint {http://arxiv.org/abs/1905.00869} {arXiv:1905.00869 [gr-qc]} \BibitemShut {NoStop}%
\bibitem [{\citenamefont {Aso}\ \emph {et~al.}(2013)\citenamefont {Aso}, \citenamefont {Michimura}, \citenamefont {Somiya}, \citenamefont {Ando}, \citenamefont {Miyakawa}, \citenamefont {Sekiguchi}, \citenamefont {Tatsumi},\ and\ \citenamefont {Yamamoto}}]{Aso:2013eba}%
  \BibitemOpen
  \bibfield  {author} {\bibinfo {author} {\bibfnamefont {Y.}~\bibnamefont {Aso}}, \bibinfo {author} {\bibfnamefont {Y.}~\bibnamefont {Michimura}}, \bibinfo {author} {\bibfnamefont {K.}~\bibnamefont {Somiya}}, \bibinfo {author} {\bibfnamefont {M.}~\bibnamefont {Ando}}, \bibinfo {author} {\bibfnamefont {O.}~\bibnamefont {Miyakawa}}, \bibinfo {author} {\bibfnamefont {T.}~\bibnamefont {Sekiguchi}}, \bibinfo {author} {\bibfnamefont {D.}~\bibnamefont {Tatsumi}}, \ and\ \bibinfo {author} {\bibfnamefont {H.}~\bibnamefont {Yamamoto}} (\bibinfo {collaboration} {KAGRA}),\ }\href {\doibase 10.1103/PhysRevD.88.043007} {\bibfield  {journal} {\bibinfo  {journal} {Phys. Rev. D}\ }\textbf {\bibinfo {volume} {88}},\ \bibinfo {pages} {043007} (\bibinfo {year} {2013})},\ \Eprint {http://arxiv.org/abs/1306.6747} {arXiv:1306.6747 [gr-qc]} \BibitemShut {NoStop}%
\bibitem [{\citenamefont {Punturo}\ \emph {et~al.}(2010)\citenamefont {Punturo} \emph {et~al.}}]{Punturo:2010zz}%
  \BibitemOpen
  \bibfield  {author} {\bibinfo {author} {\bibfnamefont {M.}~\bibnamefont {Punturo}} \emph {et~al.},\ }\href {\doibase 10.1088/0264-9381/27/19/194002} {\bibfield  {journal} {\bibinfo  {journal} {Class. Quant. Grav.}\ }\textbf {\bibinfo {volume} {27}},\ \bibinfo {pages} {194002} (\bibinfo {year} {2010})}\BibitemShut {NoStop}%
\bibitem [{\citenamefont {Reitze}\ \emph {et~al.}(2019)\citenamefont {Reitze} \emph {et~al.}}]{Reitze:2019iox}%
  \BibitemOpen
  \bibfield  {author} {\bibinfo {author} {\bibfnamefont {D.}~\bibnamefont {Reitze}} \emph {et~al.},\ }\href@noop {} {\bibfield  {journal} {\bibinfo  {journal} {Bull. Am. Astron. Soc.}\ }\textbf {\bibinfo {volume} {51}},\ \bibinfo {pages} {035} (\bibinfo {year} {2019})},\ \Eprint {http://arxiv.org/abs/1907.04833} {arXiv:1907.04833 [astro-ph.IM]} \BibitemShut {NoStop}%
\bibitem [{\citenamefont {Luo}\ \emph {et~al.}(2016)\citenamefont {Luo} \emph {et~al.}}]{TianQin:2015yph}%
  \BibitemOpen
  \bibfield  {author} {\bibinfo {author} {\bibfnamefont {J.}~\bibnamefont {Luo}} \emph {et~al.} (\bibinfo {collaboration} {TianQin}),\ }\href {\doibase 10.1088/0264-9381/33/3/035010} {\bibfield  {journal} {\bibinfo  {journal} {Class. Quant. Grav.}\ }\textbf {\bibinfo {volume} {33}},\ \bibinfo {pages} {035010} (\bibinfo {year} {2016})},\ \Eprint {http://arxiv.org/abs/1512.02076} {arXiv:1512.02076 [astro-ph.IM]} \BibitemShut {NoStop}%
\bibitem [{\citenamefont {Mei}\ \emph {et~al.}(2021)\citenamefont {Mei} \emph {et~al.}}]{TianQin:2020hid}%
  \BibitemOpen
  \bibfield  {author} {\bibinfo {author} {\bibfnamefont {J.}~\bibnamefont {Mei}} \emph {et~al.} (\bibinfo {collaboration} {TianQin}),\ }\href {\doibase 10.1093/ptep/ptaa114} {\bibfield  {journal} {\bibinfo  {journal} {PTEP}\ }\textbf {\bibinfo {volume} {2021}},\ \bibinfo {pages} {05A107} (\bibinfo {year} {2021})},\ \Eprint {http://arxiv.org/abs/2008.10332} {arXiv:2008.10332 [gr-qc]} \BibitemShut {NoStop}%
\bibitem [{\citenamefont {Amaro-Seoane}\ \emph {et~al.}(2017)\citenamefont {Amaro-Seoane} \emph {et~al.}}]{LISA:2017pwj}%
  \BibitemOpen
  \bibfield  {author} {\bibinfo {author} {\bibfnamefont {P.}~\bibnamefont {Amaro-Seoane}} \emph {et~al.} (\bibinfo {collaboration} {LISA}),\ }\href@noop {} {\  (\bibinfo {year} {2017})},\ \Eprint {http://arxiv.org/abs/1702.00786} {arXiv:1702.00786 [astro-ph.IM]} \BibitemShut {NoStop}%
\bibitem [{\citenamefont {Hu}\ and\ \citenamefont {Wu}(2017)}]{Hu:2017mde}%
  \BibitemOpen
  \bibfield  {author} {\bibinfo {author} {\bibfnamefont {W.-R.}\ \bibnamefont {Hu}}\ and\ \bibinfo {author} {\bibfnamefont {Y.-L.}\ \bibnamefont {Wu}},\ }\href {\doibase 10.1093/nsr/nwx116} {\bibfield  {journal} {\bibinfo  {journal} {Natl. Sci. Rev.}\ }\textbf {\bibinfo {volume} {4}},\ \bibinfo {pages} {685} (\bibinfo {year} {2017})}\BibitemShut {NoStop}%
\bibitem [{\citenamefont {Liu}\ \emph {et~al.}(2020)\citenamefont {Liu}, \citenamefont {Hu}, \citenamefont {Zhang},\ and\ \citenamefont {Mei}}]{Liu:2020eko}%
  \BibitemOpen
  \bibfield  {author} {\bibinfo {author} {\bibfnamefont {S.}~\bibnamefont {Liu}}, \bibinfo {author} {\bibfnamefont {Y.-M.}\ \bibnamefont {Hu}}, \bibinfo {author} {\bibfnamefont {J.-d.}\ \bibnamefont {Zhang}}, \ and\ \bibinfo {author} {\bibfnamefont {J.}~\bibnamefont {Mei}},\ }\href {\doibase 10.1103/PhysRevD.101.103027} {\bibfield  {journal} {\bibinfo  {journal} {Phys. Rev. D}\ }\textbf {\bibinfo {volume} {101}},\ \bibinfo {pages} {103027} (\bibinfo {year} {2020})},\ \Eprint {http://arxiv.org/abs/2004.14242} {arXiv:2004.14242 [astro-ph.HE]} \BibitemShut {NoStop}%
\bibitem [{\citenamefont {Liu}\ \emph {et~al.}(2022)\citenamefont {Liu}, \citenamefont {Zhu}, \citenamefont {Hu}, \citenamefont {Zhang},\ and\ \citenamefont {Ji}}]{Liu:2021yoy}%
  \BibitemOpen
  \bibfield  {author} {\bibinfo {author} {\bibfnamefont {S.}~\bibnamefont {Liu}}, \bibinfo {author} {\bibfnamefont {L.-G.}\ \bibnamefont {Zhu}}, \bibinfo {author} {\bibfnamefont {Y.-M.}\ \bibnamefont {Hu}}, \bibinfo {author} {\bibfnamefont {J.-d.}\ \bibnamefont {Zhang}}, \ and\ \bibinfo {author} {\bibfnamefont {M.-J.}\ \bibnamefont {Ji}},\ }\href {\doibase 10.1103/PhysRevD.105.023019} {\bibfield  {journal} {\bibinfo  {journal} {Phys. Rev. D}\ }\textbf {\bibinfo {volume} {105}},\ \bibinfo {pages} {023019} (\bibinfo {year} {2022})},\ \Eprint {http://arxiv.org/abs/2110.05248} {arXiv:2110.05248 [astro-ph.HE]} \BibitemShut {NoStop}%
\bibitem [{\citenamefont {Fan}\ \emph {et~al.}(2020)\citenamefont {Fan}, \citenamefont {Hu}, \citenamefont {Barausse}, \citenamefont {Sesana}, \citenamefont {Zhang}, \citenamefont {Zhang}, \citenamefont {Zi},\ and\ \citenamefont {Mei}}]{Fan:2020zhy}%
  \BibitemOpen
  \bibfield  {author} {\bibinfo {author} {\bibfnamefont {H.-M.}\ \bibnamefont {Fan}}, \bibinfo {author} {\bibfnamefont {Y.-M.}\ \bibnamefont {Hu}}, \bibinfo {author} {\bibfnamefont {E.}~\bibnamefont {Barausse}}, \bibinfo {author} {\bibfnamefont {A.}~\bibnamefont {Sesana}}, \bibinfo {author} {\bibfnamefont {J.-d.}\ \bibnamefont {Zhang}}, \bibinfo {author} {\bibfnamefont {X.}~\bibnamefont {Zhang}}, \bibinfo {author} {\bibfnamefont {T.-G.}\ \bibnamefont {Zi}}, \ and\ \bibinfo {author} {\bibfnamefont {J.}~\bibnamefont {Mei}},\ }\href {\doibase 10.1103/PhysRevD.102.063016} {\bibfield  {journal} {\bibinfo  {journal} {Phys. Rev. D}\ }\textbf {\bibinfo {volume} {102}},\ \bibinfo {pages} {063016} (\bibinfo {year} {2020})},\ \Eprint {http://arxiv.org/abs/2005.08212} {arXiv:2005.08212 [astro-ph.HE]} \BibitemShut {NoStop}%
\bibitem [{\citenamefont {Liang}\ \emph {et~al.}(2022)\citenamefont {Liang}, \citenamefont {Hu}, \citenamefont {Jiang}, \citenamefont {Cheng}, \citenamefont {Zhang},\ and\ \citenamefont {Mei}}]{Liang:2021bde}%
  \BibitemOpen
  \bibfield  {author} {\bibinfo {author} {\bibfnamefont {Z.-C.}\ \bibnamefont {Liang}}, \bibinfo {author} {\bibfnamefont {Y.-M.}\ \bibnamefont {Hu}}, \bibinfo {author} {\bibfnamefont {Y.}~\bibnamefont {Jiang}}, \bibinfo {author} {\bibfnamefont {J.}~\bibnamefont {Cheng}}, \bibinfo {author} {\bibfnamefont {J.-d.}\ \bibnamefont {Zhang}}, \ and\ \bibinfo {author} {\bibfnamefont {J.}~\bibnamefont {Mei}},\ }\href {\doibase 10.1103/PhysRevD.105.022001} {\bibfield  {journal} {\bibinfo  {journal} {Phys. Rev. D}\ }\textbf {\bibinfo {volume} {105}},\ \bibinfo {pages} {022001} (\bibinfo {year} {2022})},\ \Eprint {http://arxiv.org/abs/2107.08643} {arXiv:2107.08643 [astro-ph.CO]} \BibitemShut {NoStop}%
\bibitem [{\citenamefont {Cheng}\ \emph {et~al.}(2022)\citenamefont {Cheng}, \citenamefont {Li}, \citenamefont {Hu}, \citenamefont {Liang}, \citenamefont {Zhang},\ and\ \citenamefont {Mei}}]{Cheng:2022vct}%
  \BibitemOpen
  \bibfield  {author} {\bibinfo {author} {\bibfnamefont {J.}~\bibnamefont {Cheng}}, \bibinfo {author} {\bibfnamefont {E.-K.}\ \bibnamefont {Li}}, \bibinfo {author} {\bibfnamefont {Y.-M.}\ \bibnamefont {Hu}}, \bibinfo {author} {\bibfnamefont {Z.-C.}\ \bibnamefont {Liang}}, \bibinfo {author} {\bibfnamefont {J.-d.}\ \bibnamefont {Zhang}}, \ and\ \bibinfo {author} {\bibfnamefont {J.}~\bibnamefont {Mei}},\ }\href {\doibase 10.1103/PhysRevD.106.124027} {\bibfield  {journal} {\bibinfo  {journal} {Phys. Rev. D}\ }\textbf {\bibinfo {volume} {106}},\ \bibinfo {pages} {124027} (\bibinfo {year} {2022})},\ \Eprint {http://arxiv.org/abs/2208.11615} {arXiv:2208.11615 [gr-qc]} \BibitemShut {NoStop}%
\bibitem [{\citenamefont {Huang}\ \emph {et~al.}(2020)\citenamefont {Huang}, \citenamefont {Hu}, \citenamefont {Korol}, \citenamefont {Li}, \citenamefont {Liang}, \citenamefont {Lu}, \citenamefont {Wang}, \citenamefont {Yu},\ and\ \citenamefont {Mei}}]{Huang:2020rjf}%
  \BibitemOpen
  \bibfield  {author} {\bibinfo {author} {\bibfnamefont {S.-J.}\ \bibnamefont {Huang}}, \bibinfo {author} {\bibfnamefont {Y.-M.}\ \bibnamefont {Hu}}, \bibinfo {author} {\bibfnamefont {V.}~\bibnamefont {Korol}}, \bibinfo {author} {\bibfnamefont {P.-C.}\ \bibnamefont {Li}}, \bibinfo {author} {\bibfnamefont {Z.-C.}\ \bibnamefont {Liang}}, \bibinfo {author} {\bibfnamefont {Y.}~\bibnamefont {Lu}}, \bibinfo {author} {\bibfnamefont {H.-T.}\ \bibnamefont {Wang}}, \bibinfo {author} {\bibfnamefont {S.}~\bibnamefont {Yu}}, \ and\ \bibinfo {author} {\bibfnamefont {J.}~\bibnamefont {Mei}},\ }\href {\doibase 10.1103/PhysRevD.102.063021} {\bibfield  {journal} {\bibinfo  {journal} {Phys. Rev. D}\ }\textbf {\bibinfo {volume} {102}},\ \bibinfo {pages} {063021} (\bibinfo {year} {2020})},\ \Eprint {http://arxiv.org/abs/2005.07889} {arXiv:2005.07889 [astro-ph.HE]} \BibitemShut {NoStop}%
\bibitem [{\citenamefont {Zhu}\ \emph {et~al.}(2022)\citenamefont {Zhu}, \citenamefont {Hu}, \citenamefont {Wang}, \citenamefont {Zhang}, \citenamefont {Li}, \citenamefont {Hendry},\ and\ \citenamefont {Mei}}]{Zhu:2021aat}%
  \BibitemOpen
  \bibfield  {author} {\bibinfo {author} {\bibfnamefont {L.-G.}\ \bibnamefont {Zhu}}, \bibinfo {author} {\bibfnamefont {Y.-M.}\ \bibnamefont {Hu}}, \bibinfo {author} {\bibfnamefont {H.-T.}\ \bibnamefont {Wang}}, \bibinfo {author} {\bibfnamefont {J.-d.}\ \bibnamefont {Zhang}}, \bibinfo {author} {\bibfnamefont {X.-D.}\ \bibnamefont {Li}}, \bibinfo {author} {\bibfnamefont {M.}~\bibnamefont {Hendry}}, \ and\ \bibinfo {author} {\bibfnamefont {J.}~\bibnamefont {Mei}},\ }\href {\doibase 10.1103/PhysRevResearch.4.013247} {\bibfield  {journal} {\bibinfo  {journal} {Phys. Rev. Res.}\ }\textbf {\bibinfo {volume} {4}},\ \bibinfo {pages} {013247} (\bibinfo {year} {2022})},\ \Eprint {http://arxiv.org/abs/2104.11956} {arXiv:2104.11956 [astro-ph.CO]} \BibitemShut {NoStop}%
\bibitem [{\citenamefont {Lin}\ \emph {et~al.}(2023)\citenamefont {Lin}, \citenamefont {Zhang}, \citenamefont {Dai}, \citenamefont {Huang},\ and\ \citenamefont {Mei}}]{Lin:2023ccz}%
  \BibitemOpen
  \bibfield  {author} {\bibinfo {author} {\bibfnamefont {X.-y.}\ \bibnamefont {Lin}}, \bibinfo {author} {\bibfnamefont {J.-d.}\ \bibnamefont {Zhang}}, \bibinfo {author} {\bibfnamefont {L.}~\bibnamefont {Dai}}, \bibinfo {author} {\bibfnamefont {S.-J.}\ \bibnamefont {Huang}}, \ and\ \bibinfo {author} {\bibfnamefont {J.}~\bibnamefont {Mei}},\ }\href {\doibase 10.1103/PhysRevD.108.064020} {\bibfield  {journal} {\bibinfo  {journal} {Phys. Rev. D}\ }\textbf {\bibinfo {volume} {108}},\ \bibinfo {pages} {064020} (\bibinfo {year} {2023})},\ \Eprint {http://arxiv.org/abs/2304.04800} {arXiv:2304.04800 [gr-qc]} \BibitemShut {NoStop}%
\bibitem [{\citenamefont {Huang}\ \emph {et~al.}(2023)\citenamefont {Huang}, \citenamefont {Hu}, \citenamefont {Chen}, \citenamefont {Zhang}, \citenamefont {Li}, \citenamefont {Gao},\ and\ \citenamefont {Lin}}]{Huang:2023prq}%
  \BibitemOpen
  \bibfield  {author} {\bibinfo {author} {\bibfnamefont {S.-J.}\ \bibnamefont {Huang}}, \bibinfo {author} {\bibfnamefont {Y.-M.}\ \bibnamefont {Hu}}, \bibinfo {author} {\bibfnamefont {X.}~\bibnamefont {Chen}}, \bibinfo {author} {\bibfnamefont {J.-d.}\ \bibnamefont {Zhang}}, \bibinfo {author} {\bibfnamefont {E.-K.}\ \bibnamefont {Li}}, \bibinfo {author} {\bibfnamefont {Z.}~\bibnamefont {Gao}}, \ and\ \bibinfo {author} {\bibfnamefont {X.-Y.}\ \bibnamefont {Lin}},\ }\href {\doibase 10.1088/1475-7516/2023/08/003} {\bibfield  {journal} {\bibinfo  {journal} {JCAP}\ }\textbf {\bibinfo {volume} {08}},\ \bibinfo {pages} {003} (\bibinfo {year} {2023})},\ \Eprint {http://arxiv.org/abs/2304.10435} {arXiv:2304.10435 [astro-ph.CO]} \BibitemShut {NoStop}%
\bibitem [{\citenamefont {Shi}\ \emph {et~al.}(2019)\citenamefont {Shi}, \citenamefont {Bao}, \citenamefont {Wang}, \citenamefont {Zhang}, \citenamefont {Hu}, \citenamefont {Sesana}, \citenamefont {Barausse}, \citenamefont {Mei},\ and\ \citenamefont {Luo}}]{Shi:2019hqa}%
  \BibitemOpen
  \bibfield  {author} {\bibinfo {author} {\bibfnamefont {C.}~\bibnamefont {Shi}}, \bibinfo {author} {\bibfnamefont {J.}~\bibnamefont {Bao}}, \bibinfo {author} {\bibfnamefont {H.}~\bibnamefont {Wang}}, \bibinfo {author} {\bibfnamefont {J.-d.}\ \bibnamefont {Zhang}}, \bibinfo {author} {\bibfnamefont {Y.}~\bibnamefont {Hu}}, \bibinfo {author} {\bibfnamefont {A.}~\bibnamefont {Sesana}}, \bibinfo {author} {\bibfnamefont {E.}~\bibnamefont {Barausse}}, \bibinfo {author} {\bibfnamefont {J.}~\bibnamefont {Mei}}, \ and\ \bibinfo {author} {\bibfnamefont {J.}~\bibnamefont {Luo}},\ }\href {\doibase 10.1103/PhysRevD.100.044036} {\bibfield  {journal} {\bibinfo  {journal} {Phys. Rev. D}\ }\textbf {\bibinfo {volume} {100}},\ \bibinfo {pages} {044036} (\bibinfo {year} {2019})},\ \Eprint {http://arxiv.org/abs/1902.08922} {arXiv:1902.08922 [gr-qc]} \BibitemShut {NoStop}%
\bibitem [{\citenamefont {Bao}\ \emph {et~al.}(2019)\citenamefont {Bao}, \citenamefont {Shi}, \citenamefont {Wang}, \citenamefont {Zhang}, \citenamefont {Hu}, \citenamefont {Mei},\ and\ \citenamefont {Luo}}]{Bao:2019kgt}%
  \BibitemOpen
  \bibfield  {author} {\bibinfo {author} {\bibfnamefont {J.}~\bibnamefont {Bao}}, \bibinfo {author} {\bibfnamefont {C.}~\bibnamefont {Shi}}, \bibinfo {author} {\bibfnamefont {H.}~\bibnamefont {Wang}}, \bibinfo {author} {\bibfnamefont {J.-d.}\ \bibnamefont {Zhang}}, \bibinfo {author} {\bibfnamefont {Y.}~\bibnamefont {Hu}}, \bibinfo {author} {\bibfnamefont {J.}~\bibnamefont {Mei}}, \ and\ \bibinfo {author} {\bibfnamefont {J.}~\bibnamefont {Luo}},\ }\href {\doibase 10.1103/PhysRevD.100.084024} {\bibfield  {journal} {\bibinfo  {journal} {Phys. Rev. D}\ }\textbf {\bibinfo {volume} {100}},\ \bibinfo {pages} {084024} (\bibinfo {year} {2019})},\ \Eprint {http://arxiv.org/abs/1905.11674} {arXiv:1905.11674 [gr-qc]} \BibitemShut {NoStop}%
\bibitem [{\citenamefont {Zi}\ \emph {et~al.}(2021)\citenamefont {Zi}, \citenamefont {Zhang}, \citenamefont {Fan}, \citenamefont {Zhang}, \citenamefont {Hu}, \citenamefont {Shi},\ and\ \citenamefont {Mei}}]{Zi:2021pdp}%
  \BibitemOpen
  \bibfield  {author} {\bibinfo {author} {\bibfnamefont {T.-G.}\ \bibnamefont {Zi}}, \bibinfo {author} {\bibfnamefont {J.-D.}\ \bibnamefont {Zhang}}, \bibinfo {author} {\bibfnamefont {H.-M.}\ \bibnamefont {Fan}}, \bibinfo {author} {\bibfnamefont {X.-T.}\ \bibnamefont {Zhang}}, \bibinfo {author} {\bibfnamefont {Y.-M.}\ \bibnamefont {Hu}}, \bibinfo {author} {\bibfnamefont {C.}~\bibnamefont {Shi}}, \ and\ \bibinfo {author} {\bibfnamefont {J.}~\bibnamefont {Mei}},\ }\href {\doibase 10.1103/PhysRevD.104.064008} {\bibfield  {journal} {\bibinfo  {journal} {Phys. Rev. D}\ }\textbf {\bibinfo {volume} {104}},\ \bibinfo {pages} {064008} (\bibinfo {year} {2021})},\ \Eprint {http://arxiv.org/abs/2104.06047} {arXiv:2104.06047 [gr-qc]} \BibitemShut {NoStop}%
\bibitem [{\citenamefont {Shi}\ \emph {et~al.}(2023)\citenamefont {Shi}, \citenamefont {Ji}, \citenamefont {Zhang},\ and\ \citenamefont {Mei}}]{Shi:2022qno}%
  \BibitemOpen
  \bibfield  {author} {\bibinfo {author} {\bibfnamefont {C.}~\bibnamefont {Shi}}, \bibinfo {author} {\bibfnamefont {M.}~\bibnamefont {Ji}}, \bibinfo {author} {\bibfnamefont {J.-d.}\ \bibnamefont {Zhang}}, \ and\ \bibinfo {author} {\bibfnamefont {J.}~\bibnamefont {Mei}},\ }\href {\doibase 10.1103/PhysRevD.108.024030} {\bibfield  {journal} {\bibinfo  {journal} {Phys. Rev. D}\ }\textbf {\bibinfo {volume} {108}},\ \bibinfo {pages} {024030} (\bibinfo {year} {2023})},\ \Eprint {http://arxiv.org/abs/2210.13006} {arXiv:2210.13006 [gr-qc]} \BibitemShut {NoStop}%
\bibitem [{\citenamefont {Zel'dovich}\ and\ \citenamefont {Polnarev}(1974)}]{zel1974}%
  \BibitemOpen
  \bibfield  {author} {\bibinfo {author} {\bibfnamefont {Y.~B.}\ \bibnamefont {Zel'dovich}}\ and\ \bibinfo {author} {\bibfnamefont {A.~G.}\ \bibnamefont {Polnarev}},\ }\href@noop {} {\bibfield  {journal} {\bibinfo  {journal} {Sov. Astron. 18}\ }\textbf {\bibinfo {volume} {17}} (\bibinfo {year} {1974})}\BibitemShut {NoStop}%
\bibitem [{\citenamefont {Braginsky}\ and\ \citenamefont {Throne}(1987)}]{throne1987}%
  \BibitemOpen
  \bibfield  {author} {\bibinfo {author} {\bibfnamefont {V.~B.}\ \bibnamefont {Braginsky}}\ and\ \bibinfo {author} {\bibfnamefont {K.~S.}\ \bibnamefont {Throne}},\ }\href@noop {} {\bibfield  {journal} {\bibinfo  {journal} {Nature (London) 327}\ }\textbf {\bibinfo {volume} {123}} (\bibinfo {year} {1987})}\BibitemShut {NoStop}%
\bibitem [{\citenamefont {Christodoulou}(1991)}]{PhysRevLett.67.1486}%
  \BibitemOpen
  \bibfield  {author} {\bibinfo {author} {\bibfnamefont {D.}~\bibnamefont {Christodoulou}},\ }\href {\doibase 10.1103/PhysRevLett.67.1486} {\bibfield  {journal} {\bibinfo  {journal} {Phys. Rev. Lett.}\ }\textbf {\bibinfo {volume} {67}},\ \bibinfo {pages} {1486} (\bibinfo {year} {1991})}\BibitemShut {NoStop}%
\bibitem [{\citenamefont {Blanchet}\ and\ \citenamefont {Damour}(1992)}]{PhysRevD.46.4304}%
  \BibitemOpen
  \bibfield  {author} {\bibinfo {author} {\bibfnamefont {L.}~\bibnamefont {Blanchet}}\ and\ \bibinfo {author} {\bibfnamefont {T.}~\bibnamefont {Damour}},\ }\href {\doibase 10.1103/PhysRevD.46.4304} {\bibfield  {journal} {\bibinfo  {journal} {Phys. Rev. D}\ }\textbf {\bibinfo {volume} {46}},\ \bibinfo {pages} {4304} (\bibinfo {year} {1992})}\BibitemShut {NoStop}%
\bibitem [{\citenamefont {Flanagan}\ and\ \citenamefont {Nichols}(2017)}]{Flanagan:2015pxa}%
  \BibitemOpen
  \bibfield  {author} {\bibinfo {author} {\bibfnamefont {E.~E.}\ \bibnamefont {Flanagan}}\ and\ \bibinfo {author} {\bibfnamefont {D.~A.}\ \bibnamefont {Nichols}},\ }\href {\doibase 10.1103/PhysRevD.95.044002} {\bibfield  {journal} {\bibinfo  {journal} {Phys. Rev. D}\ }\textbf {\bibinfo {volume} {95}},\ \bibinfo {pages} {044002} (\bibinfo {year} {2017})},\ \Eprint {http://arxiv.org/abs/1510.03386} {arXiv:1510.03386 [hep-th]} \BibitemShut {NoStop}%
\bibitem [{\citenamefont {Strominger}\ and\ \citenamefont {Zhiboedov}(2016)}]{Strominger:2014pwa}%
  \BibitemOpen
  \bibfield  {author} {\bibinfo {author} {\bibfnamefont {A.}~\bibnamefont {Strominger}}\ and\ \bibinfo {author} {\bibfnamefont {A.}~\bibnamefont {Zhiboedov}},\ }\href {\doibase 10.1007/JHEP01(2016)086} {\bibfield  {journal} {\bibinfo  {journal} {JHEP}\ }\textbf {\bibinfo {volume} {01}},\ \bibinfo {pages} {086} (\bibinfo {year} {2016})},\ \Eprint {http://arxiv.org/abs/1411.5745} {arXiv:1411.5745 [hep-th]} \BibitemShut {NoStop}%
\bibitem [{\citenamefont {Strominger}(2017)}]{Strominger:2017zoo}%
  \BibitemOpen
  \bibfield  {author} {\bibinfo {author} {\bibfnamefont {A.}~\bibnamefont {Strominger}},\ }\href@noop {} {\  (\bibinfo {year} {2017})},\ \Eprint {http://arxiv.org/abs/1703.05448} {arXiv:1703.05448 [hep-th]} \BibitemShut {NoStop}%
\bibitem [{\citenamefont {Du}\ and\ \citenamefont {Nishizawa}(2016)}]{Du:2016hww}%
  \BibitemOpen
  \bibfield  {author} {\bibinfo {author} {\bibfnamefont {S.~M.}\ \bibnamefont {Du}}\ and\ \bibinfo {author} {\bibfnamefont {A.}~\bibnamefont {Nishizawa}},\ }\href {\doibase 10.1103/PhysRevD.94.104063} {\bibfield  {journal} {\bibinfo  {journal} {Phys. Rev. D}\ }\textbf {\bibinfo {volume} {94}},\ \bibinfo {pages} {104063} (\bibinfo {year} {2016})},\ \Eprint {http://arxiv.org/abs/1609.09825} {arXiv:1609.09825 [gr-qc]} \BibitemShut {NoStop}%
\bibitem [{\citenamefont {Seraj}(2021)}]{Seraj:2021qja}%
  \BibitemOpen
  \bibfield  {author} {\bibinfo {author} {\bibfnamefont {A.}~\bibnamefont {Seraj}},\ }\href {\doibase 10.1007/JHEP05(2021)283} {\bibfield  {journal} {\bibinfo  {journal} {JHEP}\ }\textbf {\bibinfo {volume} {05}},\ \bibinfo {pages} {283} (\bibinfo {year} {2021})},\ \Eprint {http://arxiv.org/abs/2103.12185} {arXiv:2103.12185 [hep-th]} \BibitemShut {NoStop}%
\bibitem [{\citenamefont {Tahura}\ \emph {et~al.}(2021)\citenamefont {Tahura}, \citenamefont {Nichols},\ and\ \citenamefont {Yagi}}]{Tahura:2021hbk}%
  \BibitemOpen
  \bibfield  {author} {\bibinfo {author} {\bibfnamefont {S.}~\bibnamefont {Tahura}}, \bibinfo {author} {\bibfnamefont {D.~A.}\ \bibnamefont {Nichols}}, \ and\ \bibinfo {author} {\bibfnamefont {K.}~\bibnamefont {Yagi}},\ }\href {\doibase 10.1103/PhysRevD.104.104010} {\bibfield  {journal} {\bibinfo  {journal} {Phys. Rev. D}\ }\textbf {\bibinfo {volume} {104}},\ \bibinfo {pages} {104010} (\bibinfo {year} {2021})},\ \Eprint {http://arxiv.org/abs/2107.02208} {arXiv:2107.02208 [gr-qc]} \BibitemShut {NoStop}%
\bibitem [{\citenamefont {Hou}\ \emph {et~al.}(2022{\natexlab{a}})\citenamefont {Hou}, \citenamefont {Zhu},\ and\ \citenamefont {Zhu}}]{Hou:2021oxe}%
  \BibitemOpen
  \bibfield  {author} {\bibinfo {author} {\bibfnamefont {S.}~\bibnamefont {Hou}}, \bibinfo {author} {\bibfnamefont {T.}~\bibnamefont {Zhu}}, \ and\ \bibinfo {author} {\bibfnamefont {Z.-H.}\ \bibnamefont {Zhu}},\ }\href {\doibase 10.1103/PhysRevD.105.024025} {\bibfield  {journal} {\bibinfo  {journal} {Phys. Rev. D}\ }\textbf {\bibinfo {volume} {105}},\ \bibinfo {pages} {024025} (\bibinfo {year} {2022}{\natexlab{a}})},\ \Eprint {http://arxiv.org/abs/2109.04238} {arXiv:2109.04238 [gr-qc]} \BibitemShut {NoStop}%
\bibitem [{\citenamefont {Hou}\ \emph {et~al.}(2022{\natexlab{b}})\citenamefont {Hou}, \citenamefont {Zhu},\ and\ \citenamefont {Zhu}}]{Hou:2021bxz}%
  \BibitemOpen
  \bibfield  {author} {\bibinfo {author} {\bibfnamefont {S.}~\bibnamefont {Hou}}, \bibinfo {author} {\bibfnamefont {T.}~\bibnamefont {Zhu}}, \ and\ \bibinfo {author} {\bibfnamefont {Z.-H.}\ \bibnamefont {Zhu}},\ }\href {\doibase 10.1088/1475-7516/2022/04/032} {\bibfield  {journal} {\bibinfo  {journal} {JCAP}\ }\textbf {\bibinfo {volume} {04}},\ \bibinfo {pages} {032} (\bibinfo {year} {2022}{\natexlab{b}})},\ \Eprint {http://arxiv.org/abs/2112.13049} {arXiv:2112.13049 [gr-qc]} \BibitemShut {NoStop}%
\bibitem [{\citenamefont {Hou}\ \emph {et~al.}(2023)\citenamefont {Hou}, \citenamefont {Wang},\ and\ \citenamefont {Zhu}}]{Hou:2023pfz}%
  \BibitemOpen
  \bibfield  {author} {\bibinfo {author} {\bibfnamefont {S.}~\bibnamefont {Hou}}, \bibinfo {author} {\bibfnamefont {A.}~\bibnamefont {Wang}}, \ and\ \bibinfo {author} {\bibfnamefont {Z.-H.}\ \bibnamefont {Zhu}},\ }\href@noop {} {\  (\bibinfo {year} {2023})},\ \Eprint {http://arxiv.org/abs/2309.01165} {arXiv:2309.01165 [gr-qc]} \BibitemShut {NoStop}%
\bibitem [{\citenamefont {Ashtekar}\ \emph {et~al.}(2020{\natexlab{a}})\citenamefont {Ashtekar}, \citenamefont {De~Lorenzo},\ and\ \citenamefont {Khera}}]{Ashtekar:2019viz}%
  \BibitemOpen
  \bibfield  {author} {\bibinfo {author} {\bibfnamefont {A.}~\bibnamefont {Ashtekar}}, \bibinfo {author} {\bibfnamefont {T.}~\bibnamefont {De~Lorenzo}}, \ and\ \bibinfo {author} {\bibfnamefont {N.}~\bibnamefont {Khera}},\ }\href {\doibase 10.1007/s10714-020-02764-1} {\bibfield  {journal} {\bibinfo  {journal} {Gen. Rel. Grav.}\ }\textbf {\bibinfo {volume} {52}},\ \bibinfo {pages} {107} (\bibinfo {year} {2020}{\natexlab{a}})},\ \Eprint {http://arxiv.org/abs/1906.00913} {arXiv:1906.00913 [gr-qc]} \BibitemShut {NoStop}%
\bibitem [{\citenamefont {Seto}(2009)}]{10.1111/j.1745-3933.2009.00758.x}%
  \BibitemOpen
  \bibfield  {author} {\bibinfo {author} {\bibfnamefont {N.}~\bibnamefont {Seto}},\ }\href {\doibase 10.1111/j.1745-3933.2009.00758.x} {\bibfield  {journal} {\bibinfo  {journal} {Mon. Not. Roy. Astron. Soc.}\ }\textbf {\bibinfo {volume} {400}},\ \bibinfo {pages} {L38} (\bibinfo {year} {2009})},\ \Eprint {http://arxiv.org/abs/0909.1379} {arXiv:0909.1379 [astro-ph.CO]} \BibitemShut {NoStop}%
\bibitem [{\citenamefont {van Haasteren}\ and\ \citenamefont {Levin}(2010)}]{vanHaasteren:2009fy}%
  \BibitemOpen
  \bibfield  {author} {\bibinfo {author} {\bibfnamefont {R.}~\bibnamefont {van Haasteren}}\ and\ \bibinfo {author} {\bibfnamefont {Y.}~\bibnamefont {Levin}},\ }\href {\doibase 10.1111/j.1365-2966.2009.15885.x} {\bibfield  {journal} {\bibinfo  {journal} {Mon. Not. Roy. Astron. Soc.}\ }\textbf {\bibinfo {volume} {401}},\ \bibinfo {pages} {2372} (\bibinfo {year} {2010})},\ \Eprint {http://arxiv.org/abs/0909.0954} {arXiv:0909.0954 [astro-ph.IM]} \BibitemShut {NoStop}%
\bibitem [{\citenamefont {Pshirkov}\ \emph {et~al.}(2010)\citenamefont {Pshirkov}, \citenamefont {Baskaran},\ and\ \citenamefont {Postnov}}]{Pshirkov:2009ak}%
  \BibitemOpen
  \bibfield  {author} {\bibinfo {author} {\bibfnamefont {M.~S.}\ \bibnamefont {Pshirkov}}, \bibinfo {author} {\bibfnamefont {D.}~\bibnamefont {Baskaran}}, \ and\ \bibinfo {author} {\bibfnamefont {K.~A.}\ \bibnamefont {Postnov}},\ }\href {\doibase 10.1111/j.1365-2966.2009.15887.x} {\bibfield  {journal} {\bibinfo  {journal} {Mon. Not. Roy. Astron. Soc.}\ }\textbf {\bibinfo {volume} {402}},\ \bibinfo {pages} {417} (\bibinfo {year} {2010})},\ \Eprint {http://arxiv.org/abs/0909.0742} {arXiv:0909.0742 [astro-ph.CO]} \BibitemShut {NoStop}%
\bibitem [{\citenamefont {Cordes}\ and\ \citenamefont {Jenet}(2012)}]{Cordes:2012zz}%
  \BibitemOpen
  \bibfield  {author} {\bibinfo {author} {\bibfnamefont {J.~M.}\ \bibnamefont {Cordes}}\ and\ \bibinfo {author} {\bibfnamefont {F.~A.}\ \bibnamefont {Jenet}},\ }\href {\doibase 10.1088/0004-637X/752/1/54} {\bibfield  {journal} {\bibinfo  {journal} {Astrophys. J.}\ }\textbf {\bibinfo {volume} {752}},\ \bibinfo {pages} {54} (\bibinfo {year} {2012})}\BibitemShut {NoStop}%
\bibitem [{\citenamefont {Madison}\ \emph {et~al.}(2014)\citenamefont {Madison}, \citenamefont {Cordes},\ and\ \citenamefont {Chatterjee}}]{madison}%
  \BibitemOpen
  \bibfield  {author} {\bibinfo {author} {\bibfnamefont {D.~R.}\ \bibnamefont {Madison}}, \bibinfo {author} {\bibfnamefont {J.~M.}\ \bibnamefont {Cordes}}, \ and\ \bibinfo {author} {\bibfnamefont {S.}~\bibnamefont {Chatterjee}},\ }\href {\doibase 10.1088/0004-637X/752/1/54} {\bibfield  {journal} {\bibinfo  {journal} {Astrophys. J.}\ }\textbf {\bibinfo {volume} {788}},\ \bibinfo {pages} {741} (\bibinfo {year} {2014})}\BibitemShut {NoStop}%
\bibitem [{\citenamefont {Arzoumanian}\ \emph {et~al.}(2015)\citenamefont {Arzoumanian} \emph {et~al.}}]{NANOGrav:2015xuc}%
  \BibitemOpen
  \bibfield  {author} {\bibinfo {author} {\bibfnamefont {Z.}~\bibnamefont {Arzoumanian}} \emph {et~al.} (\bibinfo {collaboration} {NANOGrav}),\ }\href {\doibase 10.1088/0004-637X/810/2/150} {\bibfield  {journal} {\bibinfo  {journal} {Astrophys. J.}\ }\textbf {\bibinfo {volume} {810}},\ \bibinfo {pages} {150} (\bibinfo {year} {2015})},\ \Eprint {http://arxiv.org/abs/1501.05343} {arXiv:1501.05343 [astro-ph.GA]} \BibitemShut {NoStop}%
\bibitem [{\citenamefont {Lasky}\ \emph {et~al.}(2016)\citenamefont {Lasky}, \citenamefont {Thrane}, \citenamefont {Levin}, \citenamefont {Blackman},\ and\ \citenamefont {Chen}}]{Lasky:2016knh}%
  \BibitemOpen
  \bibfield  {author} {\bibinfo {author} {\bibfnamefont {P.~D.}\ \bibnamefont {Lasky}}, \bibinfo {author} {\bibfnamefont {E.}~\bibnamefont {Thrane}}, \bibinfo {author} {\bibfnamefont {Y.}~\bibnamefont {Levin}}, \bibinfo {author} {\bibfnamefont {J.}~\bibnamefont {Blackman}}, \ and\ \bibinfo {author} {\bibfnamefont {Y.}~\bibnamefont {Chen}},\ }\href {\doibase 10.1103/PhysRevLett.117.061102} {\bibfield  {journal} {\bibinfo  {journal} {Phys. Rev. Lett.}\ }\textbf {\bibinfo {volume} {117}},\ \bibinfo {pages} {061102} (\bibinfo {year} {2016})},\ \Eprint {http://arxiv.org/abs/1605.01415} {arXiv:1605.01415 [astro-ph.HE]} \BibitemShut {NoStop}%
\bibitem [{\citenamefont {McNeill}\ \emph {et~al.}(2017)\citenamefont {McNeill}, \citenamefont {Thrane},\ and\ \citenamefont {Lasky}}]{McNeill:2017uvq}%
  \BibitemOpen
  \bibfield  {author} {\bibinfo {author} {\bibfnamefont {L.~O.}\ \bibnamefont {McNeill}}, \bibinfo {author} {\bibfnamefont {E.}~\bibnamefont {Thrane}}, \ and\ \bibinfo {author} {\bibfnamefont {P.~D.}\ \bibnamefont {Lasky}},\ }\href {\doibase 10.1103/PhysRevLett.118.181103} {\bibfield  {journal} {\bibinfo  {journal} {Phys. Rev. Lett.}\ }\textbf {\bibinfo {volume} {118}},\ \bibinfo {pages} {181103} (\bibinfo {year} {2017})},\ \Eprint {http://arxiv.org/abs/1702.01759} {arXiv:1702.01759 [astro-ph.IM]} \BibitemShut {NoStop}%
\bibitem [{\citenamefont {Divakarla}\ \emph {et~al.}(2020)\citenamefont {Divakarla}, \citenamefont {Thrane}, \citenamefont {Lasky},\ and\ \citenamefont {Whiting}}]{Divakarla:2019zjj}%
  \BibitemOpen
  \bibfield  {author} {\bibinfo {author} {\bibfnamefont {A.~K.}\ \bibnamefont {Divakarla}}, \bibinfo {author} {\bibfnamefont {E.}~\bibnamefont {Thrane}}, \bibinfo {author} {\bibfnamefont {P.~D.}\ \bibnamefont {Lasky}}, \ and\ \bibinfo {author} {\bibfnamefont {B.~F.}\ \bibnamefont {Whiting}},\ }\href {\doibase 10.1103/PhysRevD.102.023010} {\bibfield  {journal} {\bibinfo  {journal} {Phys. Rev. D}\ }\textbf {\bibinfo {volume} {102}},\ \bibinfo {pages} {023010} (\bibinfo {year} {2020})},\ \Eprint {http://arxiv.org/abs/1911.07998} {arXiv:1911.07998 [gr-qc]} \BibitemShut {NoStop}%
\bibitem [{\citenamefont {Boersma}\ \emph {et~al.}(2020)\citenamefont {Boersma}, \citenamefont {Nichols},\ and\ \citenamefont {Schmidt}}]{Boersma:2020gxx}%
  \BibitemOpen
  \bibfield  {author} {\bibinfo {author} {\bibfnamefont {O.~M.}\ \bibnamefont {Boersma}}, \bibinfo {author} {\bibfnamefont {D.~A.}\ \bibnamefont {Nichols}}, \ and\ \bibinfo {author} {\bibfnamefont {P.}~\bibnamefont {Schmidt}},\ }\href {\doibase 10.1103/PhysRevD.101.083026} {\bibfield  {journal} {\bibinfo  {journal} {Phys. Rev. D}\ }\textbf {\bibinfo {volume} {101}},\ \bibinfo {pages} {083026} (\bibinfo {year} {2020})},\ \Eprint {http://arxiv.org/abs/2002.01821} {arXiv:2002.01821 [astro-ph.HE]} \BibitemShut {NoStop}%
\bibitem [{\citenamefont {H\"ubner}\ \emph {et~al.}(2020)\citenamefont {H\"ubner}, \citenamefont {Talbot}, \citenamefont {Lasky},\ and\ \citenamefont {Thrane}}]{Hubner:2019sly}%
  \BibitemOpen
  \bibfield  {author} {\bibinfo {author} {\bibfnamefont {M.}~\bibnamefont {H\"ubner}}, \bibinfo {author} {\bibfnamefont {C.}~\bibnamefont {Talbot}}, \bibinfo {author} {\bibfnamefont {P.~D.}\ \bibnamefont {Lasky}}, \ and\ \bibinfo {author} {\bibfnamefont {E.}~\bibnamefont {Thrane}},\ }\href {\doibase 10.1103/PhysRevD.101.023011} {\bibfield  {journal} {\bibinfo  {journal} {Phys. Rev. D}\ }\textbf {\bibinfo {volume} {101}},\ \bibinfo {pages} {023011} (\bibinfo {year} {2020})},\ \Eprint {http://arxiv.org/abs/1911.12496} {arXiv:1911.12496 [astro-ph.HE]} \BibitemShut {NoStop}%
\bibitem [{\citenamefont {Khera}\ \emph {et~al.}(2021)\citenamefont {Khera}, \citenamefont {Krishnan}, \citenamefont {Ashtekar},\ and\ \citenamefont {De~Lorenzo}}]{Khera:2020mcz}%
  \BibitemOpen
  \bibfield  {author} {\bibinfo {author} {\bibfnamefont {N.}~\bibnamefont {Khera}}, \bibinfo {author} {\bibfnamefont {B.}~\bibnamefont {Krishnan}}, \bibinfo {author} {\bibfnamefont {A.}~\bibnamefont {Ashtekar}}, \ and\ \bibinfo {author} {\bibfnamefont {T.}~\bibnamefont {De~Lorenzo}},\ }\href {\doibase 10.1103/PhysRevD.103.044012} {\bibfield  {journal} {\bibinfo  {journal} {Phys. Rev. D}\ }\textbf {\bibinfo {volume} {103}},\ \bibinfo {pages} {044012} (\bibinfo {year} {2021})},\ \Eprint {http://arxiv.org/abs/2009.06351} {arXiv:2009.06351 [gr-qc]} \BibitemShut {NoStop}%
\bibitem [{\citenamefont {H\"ubner}\ \emph {et~al.}(2021)\citenamefont {H\"ubner}, \citenamefont {Lasky},\ and\ \citenamefont {Thrane}}]{Hubner:2021amk}%
  \BibitemOpen
  \bibfield  {author} {\bibinfo {author} {\bibfnamefont {M.}~\bibnamefont {H\"ubner}}, \bibinfo {author} {\bibfnamefont {P.}~\bibnamefont {Lasky}}, \ and\ \bibinfo {author} {\bibfnamefont {E.}~\bibnamefont {Thrane}},\ }\href {\doibase 10.1103/PhysRevD.104.023004} {\bibfield  {journal} {\bibinfo  {journal} {Phys. Rev. D}\ }\textbf {\bibinfo {volume} {104}},\ \bibinfo {pages} {023004} (\bibinfo {year} {2021})},\ \Eprint {http://arxiv.org/abs/2105.02879} {arXiv:2105.02879 [gr-qc]} \BibitemShut {NoStop}%
\bibitem [{\citenamefont {Zhao}\ \emph {et~al.}(2021)\citenamefont {Zhao}, \citenamefont {Liu}, \citenamefont {Cao},\ and\ \citenamefont {He}}]{Zhao:2021hmx}%
  \BibitemOpen
  \bibfield  {author} {\bibinfo {author} {\bibfnamefont {Z.-C.}\ \bibnamefont {Zhao}}, \bibinfo {author} {\bibfnamefont {X.}~\bibnamefont {Liu}}, \bibinfo {author} {\bibfnamefont {Z.}~\bibnamefont {Cao}}, \ and\ \bibinfo {author} {\bibfnamefont {X.}~\bibnamefont {He}},\ }\href {\doibase 10.1103/PhysRevD.104.064056} {\bibfield  {journal} {\bibinfo  {journal} {Phys. Rev. D}\ }\textbf {\bibinfo {volume} {104}},\ \bibinfo {pages} {064056} (\bibinfo {year} {2021})}\BibitemShut {NoStop}%
\bibitem [{\citenamefont {Agazie}\ \emph {et~al.}(2023)\citenamefont {Agazie} \emph {et~al.}}]{Agazie:2023eig}%
  \BibitemOpen
  \bibfield  {author} {\bibinfo {author} {\bibfnamefont {G.}~\bibnamefont {Agazie}} \emph {et~al.},\ }\href@noop {} {\  (\bibinfo {year} {2023})},\ \Eprint {http://arxiv.org/abs/2307.13797} {arXiv:2307.13797 [gr-qc]} \BibitemShut {NoStop}%
\bibitem [{\citenamefont {Islo}\ \emph {et~al.}(2019)\citenamefont {Islo}, \citenamefont {Simon}, \citenamefont {Burke-Spolaor},\ and\ \citenamefont {Siemens}}]{Islo:2019qht}%
  \BibitemOpen
  \bibfield  {author} {\bibinfo {author} {\bibfnamefont {K.}~\bibnamefont {Islo}}, \bibinfo {author} {\bibfnamefont {J.}~\bibnamefont {Simon}}, \bibinfo {author} {\bibfnamefont {S.}~\bibnamefont {Burke-Spolaor}}, \ and\ \bibinfo {author} {\bibfnamefont {X.}~\bibnamefont {Siemens}},\ }\href@noop {} {\  (\bibinfo {year} {2019})},\ \Eprint {http://arxiv.org/abs/1906.11936} {arXiv:1906.11936 [astro-ph.HE]} \BibitemShut {NoStop}%
\bibitem [{\citenamefont {Sun}\ \emph {et~al.}(2023)\citenamefont {Sun}, \citenamefont {Shi}, \citenamefont {Zhang},\ and\ \citenamefont {Mei}}]{Sun:2022pvh}%
  \BibitemOpen
  \bibfield  {author} {\bibinfo {author} {\bibfnamefont {S.}~\bibnamefont {Sun}}, \bibinfo {author} {\bibfnamefont {C.}~\bibnamefont {Shi}}, \bibinfo {author} {\bibfnamefont {J.-d.}\ \bibnamefont {Zhang}}, \ and\ \bibinfo {author} {\bibfnamefont {J.}~\bibnamefont {Mei}},\ }\href {\doibase 10.1103/PhysRevD.107.044023} {\bibfield  {journal} {\bibinfo  {journal} {Phys. Rev. D}\ }\textbf {\bibinfo {volume} {107}},\ \bibinfo {pages} {044023} (\bibinfo {year} {2023})},\ \Eprint {http://arxiv.org/abs/2207.13009} {arXiv:2207.13009 [gr-qc]} \BibitemShut {NoStop}%
\bibitem [{\citenamefont {Mitman}\ \emph {et~al.}(2020)\citenamefont {Mitman}, \citenamefont {Moxon}, \citenamefont {Scheel}, \citenamefont {Teukolsky}, \citenamefont {Boyle}, \citenamefont {Deppe}, \citenamefont {Kidder},\ and\ \citenamefont {Throwe}}]{Mitman:2020pbt}%
  \BibitemOpen
  \bibfield  {author} {\bibinfo {author} {\bibfnamefont {K.}~\bibnamefont {Mitman}}, \bibinfo {author} {\bibfnamefont {J.}~\bibnamefont {Moxon}}, \bibinfo {author} {\bibfnamefont {M.~A.}\ \bibnamefont {Scheel}}, \bibinfo {author} {\bibfnamefont {S.~A.}\ \bibnamefont {Teukolsky}}, \bibinfo {author} {\bibfnamefont {M.}~\bibnamefont {Boyle}}, \bibinfo {author} {\bibfnamefont {N.}~\bibnamefont {Deppe}}, \bibinfo {author} {\bibfnamefont {L.~E.}\ \bibnamefont {Kidder}}, \ and\ \bibinfo {author} {\bibfnamefont {W.}~\bibnamefont {Throwe}},\ }\href {\doibase 10.1103/PhysRevD.102.104007} {\bibfield  {journal} {\bibinfo  {journal} {Phys. Rev. D}\ }\textbf {\bibinfo {volume} {102}},\ \bibinfo {pages} {104007} (\bibinfo {year} {2020})},\ \Eprint {http://arxiv.org/abs/2007.11562} {arXiv:2007.11562 [gr-qc]} \BibitemShut {NoStop}%
\bibitem [{\citenamefont {Thorne}(1992)}]{Thorne:1992sdb}%
  \BibitemOpen
  \bibfield  {author} {\bibinfo {author} {\bibfnamefont {K.~S.}\ \bibnamefont {Thorne}},\ }\href {\doibase 10.1103/PhysRevD.45.520} {\bibfield  {journal} {\bibinfo  {journal} {Phys. Rev. D}\ }\textbf {\bibinfo {volume} {45}},\ \bibinfo {pages} {520} (\bibinfo {year} {1992})}\BibitemShut {NoStop}%
\bibitem [{\citenamefont {Favata}(2009{\natexlab{a}})}]{Favata:2008ti}%
  \BibitemOpen
  \bibfield  {author} {\bibinfo {author} {\bibfnamefont {M.}~\bibnamefont {Favata}},\ }\href {\doibase 10.1088/1742-6596/154/1/012043} {\bibfield  {journal} {\bibinfo  {journal} {J. Phys. Conf. Ser.}\ }\textbf {\bibinfo {volume} {154}},\ \bibinfo {pages} {012043} (\bibinfo {year} {2009}{\natexlab{a}})},\ \Eprint {http://arxiv.org/abs/0811.3451} {arXiv:0811.3451 [astro-ph]} \BibitemShut {NoStop}%
\bibitem [{\citenamefont {Favata}(2009{\natexlab{b}})}]{Favata:2009ii}%
  \BibitemOpen
  \bibfield  {author} {\bibinfo {author} {\bibfnamefont {M.}~\bibnamefont {Favata}},\ }\href {\doibase 10.1088/0004-637X/696/2/L159} {\bibfield  {journal} {\bibinfo  {journal} {Astrophys. J. Lett.}\ }\textbf {\bibinfo {volume} {696}},\ \bibinfo {pages} {L159} (\bibinfo {year} {2009}{\natexlab{b}})},\ \Eprint {http://arxiv.org/abs/0902.3660} {arXiv:0902.3660 [astro-ph.SR]} \BibitemShut {NoStop}%
\bibitem [{\citenamefont {Favata}(2009{\natexlab{c}})}]{Favata:2008yd}%
  \BibitemOpen
  \bibfield  {author} {\bibinfo {author} {\bibfnamefont {M.}~\bibnamefont {Favata}},\ }\href {\doibase 10.1103/PhysRevD.80.024002} {\bibfield  {journal} {\bibinfo  {journal} {Phys. Rev. D}\ }\textbf {\bibinfo {volume} {80}},\ \bibinfo {pages} {024002} (\bibinfo {year} {2009}{\natexlab{c}})},\ \Eprint {http://arxiv.org/abs/0812.0069} {arXiv:0812.0069 [gr-qc]} \BibitemShut {NoStop}%
\bibitem [{\citenamefont {Favata}(2010)}]{Favata:2010zu}%
  \BibitemOpen
  \bibfield  {author} {\bibinfo {author} {\bibfnamefont {M.}~\bibnamefont {Favata}},\ }\href {\doibase 10.1088/0264-9381/27/8/084036} {\bibfield  {journal} {\bibinfo  {journal} {Class. Quant. Grav.}\ }\textbf {\bibinfo {volume} {27}},\ \bibinfo {pages} {084036} (\bibinfo {year} {2010})},\ \Eprint {http://arxiv.org/abs/1003.3486} {arXiv:1003.3486 [gr-qc]} \BibitemShut {NoStop}%
\bibitem [{\citenamefont {Favata}(2011)}]{Favata:2011qi}%
  \BibitemOpen
  \bibfield  {author} {\bibinfo {author} {\bibfnamefont {M.}~\bibnamefont {Favata}},\ }\href {\doibase 10.1103/PhysRevD.84.124013} {\bibfield  {journal} {\bibinfo  {journal} {Phys. Rev. D}\ }\textbf {\bibinfo {volume} {84}},\ \bibinfo {pages} {124013} (\bibinfo {year} {2011})},\ \Eprint {http://arxiv.org/abs/1108.3121} {arXiv:1108.3121 [gr-qc]} \BibitemShut {NoStop}%
\bibitem [{\citenamefont {Bondi}\ \emph {et~al.}(1962)\citenamefont {Bondi}, \citenamefont {van~der Burg},\ and\ \citenamefont {Metzner}}]{Bondi:1962px}%
  \BibitemOpen
  \bibfield  {author} {\bibinfo {author} {\bibfnamefont {H.}~\bibnamefont {Bondi}}, \bibinfo {author} {\bibfnamefont {M.~G.~J.}\ \bibnamefont {van~der Burg}}, \ and\ \bibinfo {author} {\bibfnamefont {A.~W.~K.}\ \bibnamefont {Metzner}},\ }\href {\doibase 10.1098/rspa.1962.0161} {\bibfield  {journal} {\bibinfo  {journal} {Proc. Roy. Soc. Lond. A}\ }\textbf {\bibinfo {volume} {269}},\ \bibinfo {pages} {21} (\bibinfo {year} {1962})}\BibitemShut {NoStop}%
\bibitem [{\citenamefont {Sachs}(1962)}]{Sachs:1962wk}%
  \BibitemOpen
  \bibfield  {author} {\bibinfo {author} {\bibfnamefont {R.~K.}\ \bibnamefont {Sachs}},\ }\href {\doibase 10.1098/rspa.1962.0206} {\bibfield  {journal} {\bibinfo  {journal} {Proc. Roy. Soc. Lond. A}\ }\textbf {\bibinfo {volume} {270}},\ \bibinfo {pages} {103} (\bibinfo {year} {1962})}\BibitemShut {NoStop}%
\bibitem [{\citenamefont {Barnich}\ and\ \citenamefont {Troessaert}(2010)}]{Barnich:2009se}%
  \BibitemOpen
  \bibfield  {author} {\bibinfo {author} {\bibfnamefont {G.}~\bibnamefont {Barnich}}\ and\ \bibinfo {author} {\bibfnamefont {C.}~\bibnamefont {Troessaert}},\ }\href {\doibase 10.1103/PhysRevLett.105.111103} {\bibfield  {journal} {\bibinfo  {journal} {Phys. Rev. Lett.}\ }\textbf {\bibinfo {volume} {105}},\ \bibinfo {pages} {111103} (\bibinfo {year} {2010})},\ \Eprint {http://arxiv.org/abs/0909.2617} {arXiv:0909.2617 [gr-qc]} \BibitemShut {NoStop}%
\bibitem [{\citenamefont {Pasterski}\ \emph {et~al.}(2016)\citenamefont {Pasterski}, \citenamefont {Strominger},\ and\ \citenamefont {Zhiboedov}}]{Pasterski:2015tva}%
  \BibitemOpen
  \bibfield  {author} {\bibinfo {author} {\bibfnamefont {S.}~\bibnamefont {Pasterski}}, \bibinfo {author} {\bibfnamefont {A.}~\bibnamefont {Strominger}}, \ and\ \bibinfo {author} {\bibfnamefont {A.}~\bibnamefont {Zhiboedov}},\ }\href {\doibase 10.1007/JHEP12(2016)053} {\bibfield  {journal} {\bibinfo  {journal} {JHEP}\ }\textbf {\bibinfo {volume} {12}},\ \bibinfo {pages} {053} (\bibinfo {year} {2016})},\ \Eprint {http://arxiv.org/abs/1502.06120} {arXiv:1502.06120 [hep-th]} \BibitemShut {NoStop}%
\bibitem [{\citenamefont {Nichols}(2018)}]{Nichols:2018qac}%
  \BibitemOpen
  \bibfield  {author} {\bibinfo {author} {\bibfnamefont {D.~A.}\ \bibnamefont {Nichols}},\ }\href {\doibase 10.1103/PhysRevD.98.064032} {\bibfield  {journal} {\bibinfo  {journal} {Phys. Rev. D}\ }\textbf {\bibinfo {volume} {98}},\ \bibinfo {pages} {064032} (\bibinfo {year} {2018})},\ \Eprint {http://arxiv.org/abs/1807.08767} {arXiv:1807.08767 [gr-qc]} \BibitemShut {NoStop}%
\bibitem [{\citenamefont {Bieri}\ and\ \citenamefont {Garfinkle}(2014)}]{Bieri:2013ada}%
  \BibitemOpen
  \bibfield  {author} {\bibinfo {author} {\bibfnamefont {L.}~\bibnamefont {Bieri}}\ and\ \bibinfo {author} {\bibfnamefont {D.}~\bibnamefont {Garfinkle}},\ }\href {\doibase 10.1103/PhysRevD.89.084039} {\bibfield  {journal} {\bibinfo  {journal} {Phys. Rev. D}\ }\textbf {\bibinfo {volume} {89}},\ \bibinfo {pages} {084039} (\bibinfo {year} {2014})},\ \Eprint {http://arxiv.org/abs/1312.6871} {arXiv:1312.6871 [gr-qc]} \BibitemShut {NoStop}%
\bibitem [{\citenamefont {Talbot}\ \emph {et~al.}(2018)\citenamefont {Talbot}, \citenamefont {Thrane}, \citenamefont {Lasky},\ and\ \citenamefont {Lin}}]{Talbot:2018sgr}%
  \BibitemOpen
  \bibfield  {author} {\bibinfo {author} {\bibfnamefont {C.}~\bibnamefont {Talbot}}, \bibinfo {author} {\bibfnamefont {E.}~\bibnamefont {Thrane}}, \bibinfo {author} {\bibfnamefont {P.~D.}\ \bibnamefont {Lasky}}, \ and\ \bibinfo {author} {\bibfnamefont {F.}~\bibnamefont {Lin}},\ }\href {\doibase 10.1103/PhysRevD.98.064031} {\bibfield  {journal} {\bibinfo  {journal} {Phys. Rev. D}\ }\textbf {\bibinfo {volume} {98}},\ \bibinfo {pages} {064031} (\bibinfo {year} {2018})},\ \Eprint {http://arxiv.org/abs/1807.00990} {arXiv:1807.00990 [astro-ph.HE]} \BibitemShut {NoStop}%
\bibitem [{\citenamefont {Mitman}\ \emph {et~al.}(2021)\citenamefont {Mitman} \emph {et~al.}}]{Mitman:2020bjf}%
  \BibitemOpen
  \bibfield  {author} {\bibinfo {author} {\bibfnamefont {K.}~\bibnamefont {Mitman}} \emph {et~al.},\ }\href {\doibase 10.1103/PhysRevD.103.024031} {\bibfield  {journal} {\bibinfo  {journal} {Phys. Rev. D}\ }\textbf {\bibinfo {volume} {103}},\ \bibinfo {pages} {024031} (\bibinfo {year} {2021})},\ \Eprint {http://arxiv.org/abs/2011.01309} {arXiv:2011.01309 [gr-qc]} \BibitemShut {NoStop}%
\bibitem [{\citenamefont {Ezra}\ and\ \citenamefont {Roger}(1962)}]{roger1}%
  \BibitemOpen
  \bibfield  {author} {\bibinfo {author} {\bibfnamefont {N.}~\bibnamefont {Ezra}}\ and\ \bibinfo {author} {\bibfnamefont {P.}~\bibnamefont {Roger}},\ }\href {\doibase https://doi.org/10.1063/1.1724257} {\bibfield  {journal} {\bibinfo  {journal} {J. Math. Phys.}\ }\textbf {\bibinfo {volume} {3}},\ \bibinfo {pages} {566} (\bibinfo {year} {1962})},\ \Eprint {http://arxiv.org/abs/1803.01944} {arXiv:1803.01944 [astro-ph.HE]} \BibitemShut {NoStop}%
\bibitem [{\citenamefont {Ashtekar}\ \emph {et~al.}(2020{\natexlab{b}})\citenamefont {Ashtekar}, \citenamefont {De~Lorenzo},\ and\ \citenamefont {Khera}}]{Ashtekar:2019rpv}%
  \BibitemOpen
  \bibfield  {author} {\bibinfo {author} {\bibfnamefont {A.}~\bibnamefont {Ashtekar}}, \bibinfo {author} {\bibfnamefont {T.}~\bibnamefont {De~Lorenzo}}, \ and\ \bibinfo {author} {\bibfnamefont {N.}~\bibnamefont {Khera}},\ }\href {\doibase 10.1103/PhysRevD.101.044005} {\bibfield  {journal} {\bibinfo  {journal} {Phys. Rev. D}\ }\textbf {\bibinfo {volume} {101}},\ \bibinfo {pages} {044005} (\bibinfo {year} {2020}{\natexlab{b}})},\ \Eprint {http://arxiv.org/abs/1910.02907} {arXiv:1910.02907 [gr-qc]} \BibitemShut {NoStop}%
\bibitem [{\citenamefont {Armstrong}\ \emph {et~al.}(1999)\citenamefont {Armstrong}, \citenamefont {Estabrook},\ and\ \citenamefont {Tinto}}]{Armstrong_1999}%
  \BibitemOpen
  \bibfield  {author} {\bibinfo {author} {\bibfnamefont {J.~W.}\ \bibnamefont {Armstrong}}, \bibinfo {author} {\bibfnamefont {F.~B.}\ \bibnamefont {Estabrook}}, \ and\ \bibinfo {author} {\bibfnamefont {M.}~\bibnamefont {Tinto}},\ }\href {\doibase 10.1086/308110} {\bibfield  {journal} {\bibinfo  {journal} {The Astrophysical Journal}\ }\textbf {\bibinfo {volume} {527}},\ \bibinfo {pages} {814} (\bibinfo {year} {1999})}\BibitemShut {NoStop}%
\bibitem [{\citenamefont {Tinto}\ and\ \citenamefont {Armstrong}(1999)}]{Tinto:1999yr}%
  \BibitemOpen
  \bibfield  {author} {\bibinfo {author} {\bibfnamefont {M.}~\bibnamefont {Tinto}}\ and\ \bibinfo {author} {\bibfnamefont {J.~W.}\ \bibnamefont {Armstrong}},\ }\href {\doibase 10.1103/PhysRevD.59.102003} {\bibfield  {journal} {\bibinfo  {journal} {Phys. Rev. D}\ }\textbf {\bibinfo {volume} {59}},\ \bibinfo {pages} {102003} (\bibinfo {year} {1999})}\BibitemShut {NoStop}%
\bibitem [{\citenamefont {Estabrook}\ \emph {et~al.}(2000)\citenamefont {Estabrook}, \citenamefont {Tinto},\ and\ \citenamefont {Armstrong}}]{Estabrook:2000ef}%
  \BibitemOpen
  \bibfield  {author} {\bibinfo {author} {\bibfnamefont {F.~B.}\ \bibnamefont {Estabrook}}, \bibinfo {author} {\bibfnamefont {M.}~\bibnamefont {Tinto}}, \ and\ \bibinfo {author} {\bibfnamefont {J.~W.}\ \bibnamefont {Armstrong}},\ }\href {\doibase 10.1103/PhysRevD.62.042002} {\bibfield  {journal} {\bibinfo  {journal} {Phys. Rev. D}\ }\textbf {\bibinfo {volume} {62}},\ \bibinfo {pages} {042002} (\bibinfo {year} {2000})}\BibitemShut {NoStop}%
\bibitem [{\citenamefont {Dhurandhar}\ \emph {et~al.}(2002)\citenamefont {Dhurandhar}, \citenamefont {Rajesh~Nayak},\ and\ \citenamefont {Vinet}}]{Dhurandhar:2001tct}%
  \BibitemOpen
  \bibfield  {author} {\bibinfo {author} {\bibfnamefont {S.~V.}\ \bibnamefont {Dhurandhar}}, \bibinfo {author} {\bibfnamefont {K.}~\bibnamefont {Rajesh~Nayak}}, \ and\ \bibinfo {author} {\bibfnamefont {J.~Y.}\ \bibnamefont {Vinet}},\ }\href {\doibase 10.1103/PhysRevD.65.102002} {\bibfield  {journal} {\bibinfo  {journal} {Phys. Rev. D}\ }\textbf {\bibinfo {volume} {65}},\ \bibinfo {pages} {102002} (\bibinfo {year} {2002})},\ \Eprint {http://arxiv.org/abs/gr-qc/0112059} {arXiv:gr-qc/0112059} \BibitemShut {NoStop}%
\bibitem [{\citenamefont {Tinto}\ \emph {et~al.}(2004)\citenamefont {Tinto}, \citenamefont {Estabrook},\ and\ \citenamefont {Armstrong}}]{Tinto:2003vj}%
  \BibitemOpen
  \bibfield  {author} {\bibinfo {author} {\bibfnamefont {M.}~\bibnamefont {Tinto}}, \bibinfo {author} {\bibfnamefont {F.~B.}\ \bibnamefont {Estabrook}}, \ and\ \bibinfo {author} {\bibfnamefont {J.~W.}\ \bibnamefont {Armstrong}},\ }\href {\doibase 10.1103/PhysRevD.69.082001} {\bibfield  {journal} {\bibinfo  {journal} {Phys. Rev. D}\ }\textbf {\bibinfo {volume} {69}},\ \bibinfo {pages} {082001} (\bibinfo {year} {2004})},\ \Eprint {http://arxiv.org/abs/gr-qc/0310017} {arXiv:gr-qc/0310017} \BibitemShut {NoStop}%
\bibitem [{\citenamefont {Vallisneri}(2005)}]{Vallisneri:2004bn}%
  \BibitemOpen
  \bibfield  {author} {\bibinfo {author} {\bibfnamefont {M.}~\bibnamefont {Vallisneri}},\ }\href {\doibase 10.1103/PhysRevD.71.022001} {\bibfield  {journal} {\bibinfo  {journal} {Phys. Rev. D}\ }\textbf {\bibinfo {volume} {71}},\ \bibinfo {pages} {022001} (\bibinfo {year} {2005})},\ \Eprint {http://arxiv.org/abs/gr-qc/0407102} {arXiv:gr-qc/0407102} \BibitemShut {NoStop}%
\bibitem [{\citenamefont {Prince}\ \emph {et~al.}(2002)\citenamefont {Prince}, \citenamefont {Tinto}, \citenamefont {Larson},\ and\ \citenamefont {Armstrong}}]{Prince:2002hp}%
  \BibitemOpen
  \bibfield  {author} {\bibinfo {author} {\bibfnamefont {T.~A.}\ \bibnamefont {Prince}}, \bibinfo {author} {\bibfnamefont {M.}~\bibnamefont {Tinto}}, \bibinfo {author} {\bibfnamefont {S.~L.}\ \bibnamefont {Larson}}, \ and\ \bibinfo {author} {\bibfnamefont {J.~W.}\ \bibnamefont {Armstrong}},\ }\href {\doibase 10.1103/PhysRevD.66.122002} {\bibfield  {journal} {\bibinfo  {journal} {Phys. Rev. D}\ }\textbf {\bibinfo {volume} {66}},\ \bibinfo {pages} {122002} (\bibinfo {year} {2002})},\ \Eprint {http://arxiv.org/abs/gr-qc/0209039} {arXiv:gr-qc/0209039} \BibitemShut {NoStop}%
\bibitem [{\citenamefont {Hu}\ \emph {et~al.}(2018)\citenamefont {Hu}, \citenamefont {Li}, \citenamefont {Wang}, \citenamefont {Feng}, \citenamefont {Zhou}, \citenamefont {Hu}, \citenamefont {Hu}, \citenamefont {Mei},\ and\ \citenamefont {Shao}}]{Hu:2018yqb}%
  \BibitemOpen
  \bibfield  {author} {\bibinfo {author} {\bibfnamefont {X.-C.}\ \bibnamefont {Hu}}, \bibinfo {author} {\bibfnamefont {X.-H.}\ \bibnamefont {Li}}, \bibinfo {author} {\bibfnamefont {Y.}~\bibnamefont {Wang}}, \bibinfo {author} {\bibfnamefont {W.-F.}\ \bibnamefont {Feng}}, \bibinfo {author} {\bibfnamefont {M.-Y.}\ \bibnamefont {Zhou}}, \bibinfo {author} {\bibfnamefont {Y.-M.}\ \bibnamefont {Hu}}, \bibinfo {author} {\bibfnamefont {S.-C.}\ \bibnamefont {Hu}}, \bibinfo {author} {\bibfnamefont {J.-W.}\ \bibnamefont {Mei}}, \ and\ \bibinfo {author} {\bibfnamefont {C.-G.}\ \bibnamefont {Shao}},\ }\href {\doibase 10.1088/1361-6382/aab52f} {\bibfield  {journal} {\bibinfo  {journal} {Class. Quant. Grav.}\ }\textbf {\bibinfo {volume} {35}},\ \bibinfo {pages} {095008} (\bibinfo {year} {2018})},\ \Eprint {http://arxiv.org/abs/1803.03368} {arXiv:1803.03368 [gr-qc]} \BibitemShut {NoStop}%
\bibitem [{\citenamefont {Katz}\ \emph {et~al.}(2022)\citenamefont {Katz}, \citenamefont {Bayle}, \citenamefont {Chua},\ and\ \citenamefont {Vallisneri}}]{Katz:2022yqe}%
  \BibitemOpen
  \bibfield  {author} {\bibinfo {author} {\bibfnamefont {M.~L.}\ \bibnamefont {Katz}}, \bibinfo {author} {\bibfnamefont {J.-B.}\ \bibnamefont {Bayle}}, \bibinfo {author} {\bibfnamefont {A.~J.~K.}\ \bibnamefont {Chua}}, \ and\ \bibinfo {author} {\bibfnamefont {M.}~\bibnamefont {Vallisneri}},\ }\href {\doibase 10.1103/PhysRevD.106.103001} {\bibfield  {journal} {\bibinfo  {journal} {Phys. Rev. D}\ }\textbf {\bibinfo {volume} {106}},\ \bibinfo {pages} {103001} (\bibinfo {year} {2022})},\ \Eprint {http://arxiv.org/abs/2204.06633} {arXiv:2204.06633 [gr-qc]} \BibitemShut {NoStop}%
\bibitem [{\citenamefont {Krolak}\ \emph {et~al.}(2004)\citenamefont {Krolak}, \citenamefont {Tinto},\ and\ \citenamefont {Vallisneri}}]{Krolak:2004xp}%
  \BibitemOpen
  \bibfield  {author} {\bibinfo {author} {\bibfnamefont {A.}~\bibnamefont {Krolak}}, \bibinfo {author} {\bibfnamefont {M.}~\bibnamefont {Tinto}}, \ and\ \bibinfo {author} {\bibfnamefont {M.}~\bibnamefont {Vallisneri}},\ }\href {\doibase 10.1103/PhysRevD.70.022003} {\bibfield  {journal} {\bibinfo  {journal} {Phys. Rev. D}\ }\textbf {\bibinfo {volume} {70}},\ \bibinfo {pages} {022003} (\bibinfo {year} {2004})},\ \bibinfo {note} {[Erratum: Phys.Rev.D 76, 069901 (2007)]},\ \Eprint {http://arxiv.org/abs/gr-qc/0401108} {arXiv:gr-qc/0401108} \BibitemShut {NoStop}%
\bibitem [{\citenamefont {Marsat}\ \emph {et~al.}(2021)\citenamefont {Marsat}, \citenamefont {Baker},\ and\ \citenamefont {Dal~Canton}}]{Marsat:2020rtl}%
  \BibitemOpen
  \bibfield  {author} {\bibinfo {author} {\bibfnamefont {S.}~\bibnamefont {Marsat}}, \bibinfo {author} {\bibfnamefont {J.~G.}\ \bibnamefont {Baker}}, \ and\ \bibinfo {author} {\bibfnamefont {T.}~\bibnamefont {Dal~Canton}},\ }\href {\doibase 10.1103/PhysRevD.103.083011} {\bibfield  {journal} {\bibinfo  {journal} {Phys. Rev. D}\ }\textbf {\bibinfo {volume} {103}},\ \bibinfo {pages} {083011} (\bibinfo {year} {2021})},\ \Eprint {http://arxiv.org/abs/2003.00357} {arXiv:2003.00357 [gr-qc]} \BibitemShut {NoStop}%
\bibitem [{\citenamefont {Toubiana}\ \emph {et~al.}(2020)\citenamefont {Toubiana}, \citenamefont {Marsat}, \citenamefont {Babak}, \citenamefont {Baker},\ and\ \citenamefont {Dal~Canton}}]{Toubiana:2020cqv}%
  \BibitemOpen
  \bibfield  {author} {\bibinfo {author} {\bibfnamefont {A.}~\bibnamefont {Toubiana}}, \bibinfo {author} {\bibfnamefont {S.}~\bibnamefont {Marsat}}, \bibinfo {author} {\bibfnamefont {S.}~\bibnamefont {Babak}}, \bibinfo {author} {\bibfnamefont {J.}~\bibnamefont {Baker}}, \ and\ \bibinfo {author} {\bibfnamefont {T.}~\bibnamefont {Dal~Canton}},\ }\href {\doibase 10.1103/PhysRevD.102.124037} {\bibfield  {journal} {\bibinfo  {journal} {Phys. Rev. D}\ }\textbf {\bibinfo {volume} {102}},\ \bibinfo {pages} {124037} (\bibinfo {year} {2020})},\ \Eprint {http://arxiv.org/abs/2007.08544} {arXiv:2007.08544 [gr-qc]} \BibitemShut {NoStop}%
\bibitem [{\citenamefont {Lyu}\ \emph {et~al.}(2023)\citenamefont {Lyu}, \citenamefont {Li},\ and\ \citenamefont {Hu}}]{Lyu:2023ctt}%
  \BibitemOpen
  \bibfield  {author} {\bibinfo {author} {\bibfnamefont {X.}~\bibnamefont {Lyu}}, \bibinfo {author} {\bibfnamefont {E.-K.}\ \bibnamefont {Li}}, \ and\ \bibinfo {author} {\bibfnamefont {Y.-M.}\ \bibnamefont {Hu}},\ }\href {\doibase 10.1103/PhysRevD.108.083023} {\bibfield  {journal} {\bibinfo  {journal} {Phys. Rev. D}\ }\textbf {\bibinfo {volume} {108}},\ \bibinfo {pages} {083023} (\bibinfo {year} {2023})},\ \Eprint {http://arxiv.org/abs/2307.12244} {arXiv:2307.12244 [gr-qc]} \BibitemShut {NoStop}%
\bibitem [{\citenamefont {Thrane}\ and\ \citenamefont {Talbot}(2019)}]{Thrane:2018qnx}%
  \BibitemOpen
  \bibfield  {author} {\bibinfo {author} {\bibfnamefont {E.}~\bibnamefont {Thrane}}\ and\ \bibinfo {author} {\bibfnamefont {C.}~\bibnamefont {Talbot}},\ }\href {\doibase 10.1017/pasa.2019.2} {\bibfield  {journal} {\bibinfo  {journal} {Publ. Astron. Soc. Austral.}\ }\textbf {\bibinfo {volume} {36}},\ \bibinfo {pages} {e010} (\bibinfo {year} {2019})},\ \bibinfo {note} {[Erratum: Publ.Astron.Soc.Austral. 37, e036 (2020)]},\ \Eprint {http://arxiv.org/abs/1809.02293} {arXiv:1809.02293 [astro-ph.IM]} \BibitemShut {NoStop}%
\bibitem [{\citenamefont {Goncharov}\ \emph {et~al.}(2023)\citenamefont {Goncharov}, \citenamefont {Donnay},\ and\ \citenamefont {Harms}}]{Goncharov:2023woe}%
  \BibitemOpen
  \bibfield  {author} {\bibinfo {author} {\bibfnamefont {B.}~\bibnamefont {Goncharov}}, \bibinfo {author} {\bibfnamefont {L.}~\bibnamefont {Donnay}}, \ and\ \bibinfo {author} {\bibfnamefont {J.}~\bibnamefont {Harms}},\ }\href@noop {} {\  (\bibinfo {year} {2023})},\ \Eprint {http://arxiv.org/abs/2310.10718} {arXiv:2310.10718 [gr-qc]} \BibitemShut {NoStop}%
\bibitem [{\citenamefont {Speagle}(2020)}]{Speagle:2019ivv}%
  \BibitemOpen
  \bibfield  {author} {\bibinfo {author} {\bibfnamefont {J.~S.}\ \bibnamefont {Speagle}},\ }\href {\doibase 10.1093/mnras/staa278} {\bibfield  {journal} {\bibinfo  {journal} {Mon. Not. Roy. Astron. Soc.}\ }\textbf {\bibinfo {volume} {493}},\ \bibinfo {pages} {3132} (\bibinfo {year} {2020})},\ \Eprint {http://arxiv.org/abs/1904.02180} {arXiv:1904.02180 [astro-ph.IM]} \BibitemShut {NoStop}%
\bibitem [{\citenamefont {Koposov}\ \emph {et~al.}(2023)\citenamefont {Koposov}, \citenamefont {Speagle}, \citenamefont {Barbary}, \citenamefont {Ashton}, \citenamefont {Bennett}, \citenamefont {Buchner}, \citenamefont {Scheffler}, \citenamefont {Cook}, \citenamefont {Talbot}, \citenamefont {Guillochon}, \citenamefont {Cubillos}, \citenamefont {Ramos}, \citenamefont {Johnson}, \citenamefont {Lang}, \citenamefont {Ilya}, \citenamefont {Dartiailh}, \citenamefont {Nitz}, \citenamefont {McCluskey},\ and\ \citenamefont {Archibald}}]{sergey_koposov_2023_8408702}%
  \BibitemOpen
  \bibfield  {author} {\bibinfo {author} {\bibfnamefont {S.}~\bibnamefont {Koposov}}, \bibinfo {author} {\bibfnamefont {J.}~\bibnamefont {Speagle}}, \bibinfo {author} {\bibfnamefont {K.}~\bibnamefont {Barbary}}, \bibinfo {author} {\bibfnamefont {G.}~\bibnamefont {Ashton}}, \bibinfo {author} {\bibfnamefont {E.}~\bibnamefont {Bennett}}, \bibinfo {author} {\bibfnamefont {J.}~\bibnamefont {Buchner}}, \bibinfo {author} {\bibfnamefont {C.}~\bibnamefont {Scheffler}}, \bibinfo {author} {\bibfnamefont {B.}~\bibnamefont {Cook}}, \bibinfo {author} {\bibfnamefont {C.}~\bibnamefont {Talbot}}, \bibinfo {author} {\bibfnamefont {J.}~\bibnamefont {Guillochon}}, \bibinfo {author} {\bibfnamefont {P.}~\bibnamefont {Cubillos}}, \bibinfo {author} {\bibfnamefont {A.~A.}\ \bibnamefont {Ramos}}, \bibinfo {author} {\bibfnamefont {B.}~\bibnamefont {Johnson}}, \bibinfo {author} {\bibfnamefont {D.}~\bibnamefont {Lang}}, \bibinfo {author} {\bibnamefont {Ilya}}, \bibinfo {author} {\bibfnamefont {M.}~\bibnamefont {Dartiailh}}, \bibinfo
  {author} {\bibfnamefont {A.}~\bibnamefont {Nitz}}, \bibinfo {author} {\bibfnamefont {A.}~\bibnamefont {McCluskey}}, \ and\ \bibinfo {author} {\bibfnamefont {A.}~\bibnamefont {Archibald}},\ }\href {\doibase 10.5281/zenodo.8408702} {\enquote {\bibinfo {title} {joshspeagle/dynesty: v2.1.3},}\ } (\bibinfo {year} {2023})\BibitemShut {NoStop}%
\bibitem [{\citenamefont {Skilling}(2004)}]{10.1063/1.1835238}%
  \BibitemOpen
  \bibfield  {author} {\bibinfo {author} {\bibfnamefont {J.}~\bibnamefont {Skilling}},\ }\href {\doibase 10.1063/1.1835238} {\bibfield  {journal} {\bibinfo  {journal} {AIP Conference Proceedings}\ }\textbf {\bibinfo {volume} {735}},\ \bibinfo {pages} {395} (\bibinfo {year} {2004})},\ \Eprint {http://arxiv.org/abs/https://pubs.aip.org/aip/acp/article-pdf/735/1/395/11702789/395\_1\_online.pdf} {https://pubs.aip.org/aip/acp/article-pdf/735/1/395/11702789/395\_1\_online.pdf} \BibitemShut {NoStop}%
\bibitem [{\citenamefont {Skilling}(2006)}]{Skilling:2006gxv}%
  \BibitemOpen
  \bibfield  {author} {\bibinfo {author} {\bibfnamefont {J.}~\bibnamefont {Skilling}},\ }\href {\doibase 10.1214/06-BA127} {\bibfield  {journal} {\bibinfo  {journal} {Bayesian Analysis}\ }\textbf {\bibinfo {volume} {1}},\ \bibinfo {pages} {833} (\bibinfo {year} {2006})}\BibitemShut {NoStop}%
\bibitem [{\citenamefont {Pratten}\ \emph {et~al.}(2020)\citenamefont {Pratten}, \citenamefont {Husa}, \citenamefont {Garcia-Quiros}, \citenamefont {Colleoni}, \citenamefont {Ramos-Buades}, \citenamefont {Estelles},\ and\ \citenamefont {Jaume}}]{Pratten:2020fqn}%
  \BibitemOpen
  \bibfield  {author} {\bibinfo {author} {\bibfnamefont {G.}~\bibnamefont {Pratten}}, \bibinfo {author} {\bibfnamefont {S.}~\bibnamefont {Husa}}, \bibinfo {author} {\bibfnamefont {C.}~\bibnamefont {Garcia-Quiros}}, \bibinfo {author} {\bibfnamefont {M.}~\bibnamefont {Colleoni}}, \bibinfo {author} {\bibfnamefont {A.}~\bibnamefont {Ramos-Buades}}, \bibinfo {author} {\bibfnamefont {H.}~\bibnamefont {Estelles}}, \ and\ \bibinfo {author} {\bibfnamefont {R.}~\bibnamefont {Jaume}},\ }\href {\doibase 10.1103/PhysRevD.102.064001} {\bibfield  {journal} {\bibinfo  {journal} {Phys. Rev. D}\ }\textbf {\bibinfo {volume} {102}},\ \bibinfo {pages} {064001} (\bibinfo {year} {2020})},\ \Eprint {http://arxiv.org/abs/2001.11412} {arXiv:2001.11412 [gr-qc]} \BibitemShut {NoStop}%
\bibitem [{\citenamefont {Garc\'\i{}a-Quir\'os}\ \emph {et~al.}(2020)\citenamefont {Garc\'\i{}a-Quir\'os}, \citenamefont {Colleoni}, \citenamefont {Husa}, \citenamefont {Estell\'es}, \citenamefont {Pratten}, \citenamefont {Ramos-Buades}, \citenamefont {Mateu-Lucena},\ and\ \citenamefont {Jaume}}]{Garcia-Quiros:2020qpx}%
  \BibitemOpen
  \bibfield  {author} {\bibinfo {author} {\bibfnamefont {C.}~\bibnamefont {Garc\'\i{}a-Quir\'os}}, \bibinfo {author} {\bibfnamefont {M.}~\bibnamefont {Colleoni}}, \bibinfo {author} {\bibfnamefont {S.}~\bibnamefont {Husa}}, \bibinfo {author} {\bibfnamefont {H.}~\bibnamefont {Estell\'es}}, \bibinfo {author} {\bibfnamefont {G.}~\bibnamefont {Pratten}}, \bibinfo {author} {\bibfnamefont {A.}~\bibnamefont {Ramos-Buades}}, \bibinfo {author} {\bibfnamefont {M.}~\bibnamefont {Mateu-Lucena}}, \ and\ \bibinfo {author} {\bibfnamefont {R.}~\bibnamefont {Jaume}},\ }\href {\doibase 10.1103/PhysRevD.102.064002} {\bibfield  {journal} {\bibinfo  {journal} {Phys. Rev. D}\ }\textbf {\bibinfo {volume} {102}},\ \bibinfo {pages} {064002} (\bibinfo {year} {2020})},\ \Eprint {http://arxiv.org/abs/2001.10914} {arXiv:2001.10914 [gr-qc]} \BibitemShut {NoStop}%
\bibitem [{\citenamefont {Colleoni}\ \emph {et~al.}(2021)\citenamefont {Colleoni}, \citenamefont {Mateu-Lucena}, \citenamefont {Estell\'es}, \citenamefont {Garc\'\i{}a-Quir\'os}, \citenamefont {Keitel}, \citenamefont {Pratten}, \citenamefont {Ramos-Buades},\ and\ \citenamefont {Husa}}]{Colleoni:2020tgc}%
  \BibitemOpen
  \bibfield  {author} {\bibinfo {author} {\bibfnamefont {M.}~\bibnamefont {Colleoni}}, \bibinfo {author} {\bibfnamefont {M.}~\bibnamefont {Mateu-Lucena}}, \bibinfo {author} {\bibfnamefont {H.}~\bibnamefont {Estell\'es}}, \bibinfo {author} {\bibfnamefont {C.}~\bibnamefont {Garc\'\i{}a-Quir\'os}}, \bibinfo {author} {\bibfnamefont {D.}~\bibnamefont {Keitel}}, \bibinfo {author} {\bibfnamefont {G.}~\bibnamefont {Pratten}}, \bibinfo {author} {\bibfnamefont {A.}~\bibnamefont {Ramos-Buades}}, \ and\ \bibinfo {author} {\bibfnamefont {S.}~\bibnamefont {Husa}},\ }\href {\doibase 10.1103/PhysRevD.103.024029} {\bibfield  {journal} {\bibinfo  {journal} {Phys. Rev. D}\ }\textbf {\bibinfo {volume} {103}},\ \bibinfo {pages} {024029} (\bibinfo {year} {2021})},\ \Eprint {http://arxiv.org/abs/2010.05830} {arXiv:2010.05830 [gr-qc]} \BibitemShut {NoStop}%
\bibitem [{\citenamefont {{LIGO Scientific Collaboration}}(2018)}]{LALSuite}%
  \BibitemOpen
  \bibfield  {author} {\bibinfo {author} {\bibnamefont {{LIGO Scientific Collaboration}}},\ }\href {\doibase 10.7935/GT1W-FZ16} {\enquote {\bibinfo {title} {{LIGO} {A}lgorithm {L}ibrary - {LALS}uite},}\ }\bibinfo {howpublished} {free software (GPL)} (\bibinfo {year} {2018})\BibitemShut {NoStop}%
\bibitem [{\citenamefont {McKechan}\ \emph {et~al.}(2010)\citenamefont {McKechan}, \citenamefont {Robinson},\ and\ \citenamefont {Sathyaprakash}}]{McKechan:2010kp}%
  \BibitemOpen
  \bibfield  {author} {\bibinfo {author} {\bibfnamefont {D.~J.~A.}\ \bibnamefont {McKechan}}, \bibinfo {author} {\bibfnamefont {C.}~\bibnamefont {Robinson}}, \ and\ \bibinfo {author} {\bibfnamefont {B.~S.}\ \bibnamefont {Sathyaprakash}},\ }\href {\doibase 10.1088/0264-9381/27/8/084020} {\bibfield  {journal} {\bibinfo  {journal} {Class. Quant. Grav.}\ }\textbf {\bibinfo {volume} {27}},\ \bibinfo {pages} {084020} (\bibinfo {year} {2010})},\ \Eprint {http://arxiv.org/abs/1003.2939} {arXiv:1003.2939 [gr-qc]} \BibitemShut {NoStop}%
\bibitem [{\citenamefont {Wang}\ \emph {et~al.}(2019)\citenamefont {Wang} \emph {et~al.}}]{haitian}%
  \BibitemOpen
  \bibfield  {author} {\bibinfo {author} {\bibfnamefont {H.-T.}\ \bibnamefont {Wang}} \emph {et~al.},\ }\href {\doibase 10.1103/PhysRevD.100.043003} {\bibfield  {journal} {\bibinfo  {journal} {Phys. Rev. D}\ }\textbf {\bibinfo {volume} {100}},\ \bibinfo {pages} {043003} (\bibinfo {year} {2019})},\ \Eprint {http://arxiv.org/abs/1902.04423} {arXiv:1902.04423 [astro-ph.HE]} \BibitemShut {NoStop}%
\bibitem [{\citenamefont {Gasparotto}\ \emph {et~al.}(2023)\citenamefont {Gasparotto}, \citenamefont {Vicente}, \citenamefont {Blas}, \citenamefont {Jenkins},\ and\ \citenamefont {Barausse}}]{Gasparotto:2023fcg}%
  \BibitemOpen
  \bibfield  {author} {\bibinfo {author} {\bibfnamefont {S.}~\bibnamefont {Gasparotto}}, \bibinfo {author} {\bibfnamefont {R.}~\bibnamefont {Vicente}}, \bibinfo {author} {\bibfnamefont {D.}~\bibnamefont {Blas}}, \bibinfo {author} {\bibfnamefont {A.~C.}\ \bibnamefont {Jenkins}}, \ and\ \bibinfo {author} {\bibfnamefont {E.}~\bibnamefont {Barausse}},\ }\href {\doibase 10.1103/PhysRevD.107.124033} {\bibfield  {journal} {\bibinfo  {journal} {Phys. Rev. D}\ }\textbf {\bibinfo {volume} {107}},\ \bibinfo {pages} {124033} (\bibinfo {year} {2023})},\ \Eprint {http://arxiv.org/abs/2301.13228} {arXiv:2301.13228 [gr-qc]} \BibitemShut {NoStop}%
\bibitem [{\citenamefont {Vallisneri}(2008)}]{Vallisneri:2007ev}%
  \BibitemOpen
  \bibfield  {author} {\bibinfo {author} {\bibfnamefont {M.}~\bibnamefont {Vallisneri}},\ }\href {\doibase 10.1103/PhysRevD.77.042001} {\bibfield  {journal} {\bibinfo  {journal} {Phys. Rev. D}\ }\textbf {\bibinfo {volume} {77}},\ \bibinfo {pages} {042001} (\bibinfo {year} {2008})},\ \Eprint {http://arxiv.org/abs/gr-qc/0703086} {arXiv:gr-qc/0703086} \BibitemShut {NoStop}%
\bibitem [{\citenamefont {Baird}\ \emph {et~al.}(2013)\citenamefont {Baird}, \citenamefont {Fairhurst}, \citenamefont {Hannam},\ and\ \citenamefont {Murphy}}]{Baird:2012cu}%
  \BibitemOpen
  \bibfield  {author} {\bibinfo {author} {\bibfnamefont {E.}~\bibnamefont {Baird}}, \bibinfo {author} {\bibfnamefont {S.}~\bibnamefont {Fairhurst}}, \bibinfo {author} {\bibfnamefont {M.}~\bibnamefont {Hannam}}, \ and\ \bibinfo {author} {\bibfnamefont {P.}~\bibnamefont {Murphy}},\ }\href {\doibase 10.1103/PhysRevD.87.024035} {\bibfield  {journal} {\bibinfo  {journal} {Phys. Rev. D}\ }\textbf {\bibinfo {volume} {87}},\ \bibinfo {pages} {024035} (\bibinfo {year} {2013})},\ \Eprint {http://arxiv.org/abs/1211.0546} {arXiv:1211.0546 [gr-qc]} \BibitemShut {NoStop}%
\bibitem [{\citenamefont {Chatziioannou}\ \emph {et~al.}(2017)\citenamefont {Chatziioannou}, \citenamefont {Klein}, \citenamefont {Yunes},\ and\ \citenamefont {Cornish}}]{Chatziioannou:2017tdw}%
  \BibitemOpen
  \bibfield  {author} {\bibinfo {author} {\bibfnamefont {K.}~\bibnamefont {Chatziioannou}}, \bibinfo {author} {\bibfnamefont {A.}~\bibnamefont {Klein}}, \bibinfo {author} {\bibfnamefont {N.}~\bibnamefont {Yunes}}, \ and\ \bibinfo {author} {\bibfnamefont {N.}~\bibnamefont {Cornish}},\ }\href {\doibase 10.1103/PhysRevD.95.104004} {\bibfield  {journal} {\bibinfo  {journal} {Phys. Rev. D}\ }\textbf {\bibinfo {volume} {95}},\ \bibinfo {pages} {104004} (\bibinfo {year} {2017})},\ \Eprint {http://arxiv.org/abs/1703.03967} {arXiv:1703.03967 [gr-qc]} \BibitemShut {NoStop}%
\bibitem [{\citenamefont {Mangiagli}\ \emph {et~al.}(2019)\citenamefont {Mangiagli}, \citenamefont {Klein}, \citenamefont {Sesana}, \citenamefont {Barausse},\ and\ \citenamefont {Colpi}}]{Mangiagli:2018kpu}%
  \BibitemOpen
  \bibfield  {author} {\bibinfo {author} {\bibfnamefont {A.}~\bibnamefont {Mangiagli}}, \bibinfo {author} {\bibfnamefont {A.}~\bibnamefont {Klein}}, \bibinfo {author} {\bibfnamefont {A.}~\bibnamefont {Sesana}}, \bibinfo {author} {\bibfnamefont {E.}~\bibnamefont {Barausse}}, \ and\ \bibinfo {author} {\bibfnamefont {M.}~\bibnamefont {Colpi}},\ }\href {\doibase 10.1103/PhysRevD.99.064056} {\bibfield  {journal} {\bibinfo  {journal} {Phys. Rev. D}\ }\textbf {\bibinfo {volume} {99}},\ \bibinfo {pages} {064056} (\bibinfo {year} {2019})},\ \Eprint {http://arxiv.org/abs/1811.01805} {arXiv:1811.01805 [gr-qc]} \BibitemShut {NoStop}%
\bibitem [{\citenamefont {P\"urrer}\ and\ \citenamefont {Haster}(2020)}]{Purrer:2019jcp}%
  \BibitemOpen
  \bibfield  {author} {\bibinfo {author} {\bibfnamefont {M.}~\bibnamefont {P\"urrer}}\ and\ \bibinfo {author} {\bibfnamefont {C.-J.}\ \bibnamefont {Haster}},\ }\href {\doibase 10.1103/PhysRevResearch.2.023151} {\bibfield  {journal} {\bibinfo  {journal} {Phys. Rev. Res.}\ }\textbf {\bibinfo {volume} {2}},\ \bibinfo {pages} {023151} (\bibinfo {year} {2020})},\ \Eprint {http://arxiv.org/abs/1912.10055} {arXiv:1912.10055 [gr-qc]} \BibitemShut {NoStop}%
\bibitem [{\citenamefont {Toubiana}\ \emph {et~al.}(2023)\citenamefont {Toubiana}, \citenamefont {Pompili}, \citenamefont {Buonanno}, \citenamefont {Gair},\ and\ \citenamefont {Katz}}]{Toubiana:2023cwr}%
  \BibitemOpen
  \bibfield  {author} {\bibinfo {author} {\bibfnamefont {A.}~\bibnamefont {Toubiana}}, \bibinfo {author} {\bibfnamefont {L.}~\bibnamefont {Pompili}}, \bibinfo {author} {\bibfnamefont {A.}~\bibnamefont {Buonanno}}, \bibinfo {author} {\bibfnamefont {J.~R.}\ \bibnamefont {Gair}}, \ and\ \bibinfo {author} {\bibfnamefont {M.~L.}\ \bibnamefont {Katz}},\ }\href@noop {} {\  (\bibinfo {year} {2023})},\ \Eprint {http://arxiv.org/abs/2307.15086} {arXiv:2307.15086 [gr-qc]} \BibitemShut {NoStop}%
\bibitem [{\citenamefont {Pompili}\ \emph {et~al.}(2023)\citenamefont {Pompili} \emph {et~al.}}]{Pompili:2023tna}%
  \BibitemOpen
  \bibfield  {author} {\bibinfo {author} {\bibfnamefont {L.}~\bibnamefont {Pompili}} \emph {et~al.},\ }\href {\doibase 10.1103/PhysRevD.108.124035} {\bibfield  {journal} {\bibinfo  {journal} {Phys. Rev. D}\ }\textbf {\bibinfo {volume} {108}},\ \bibinfo {pages} {124035} (\bibinfo {year} {2023})},\ \Eprint {http://arxiv.org/abs/2303.18039} {arXiv:2303.18039 [gr-qc]} \BibitemShut {NoStop}%
\bibitem [{\citenamefont {Ossokine}\ \emph {et~al.}(2020)\citenamefont {Ossokine} \emph {et~al.}}]{Ossokine:2020kjp}%
  \BibitemOpen
  \bibfield  {author} {\bibinfo {author} {\bibfnamefont {S.}~\bibnamefont {Ossokine}} \emph {et~al.},\ }\href {\doibase 10.1103/PhysRevD.102.044055} {\bibfield  {journal} {\bibinfo  {journal} {Phys. Rev. D}\ }\textbf {\bibinfo {volume} {102}},\ \bibinfo {pages} {044055} (\bibinfo {year} {2020})},\ \Eprint {http://arxiv.org/abs/2004.09442} {arXiv:2004.09442 [gr-qc]} \BibitemShut {NoStop}%
\bibitem [{\citenamefont {van~der Walt}\ \emph {et~al.}(2011)\citenamefont {van~der Walt}, \citenamefont {Colbert},\ and\ \citenamefont {Varoquaux}}]{vanderWalt:2011bqk}%
  \BibitemOpen
  \bibfield  {author} {\bibinfo {author} {\bibfnamefont {S.}~\bibnamefont {van~der Walt}}, \bibinfo {author} {\bibfnamefont {S.~C.}\ \bibnamefont {Colbert}}, \ and\ \bibinfo {author} {\bibfnamefont {G.}~\bibnamefont {Varoquaux}},\ }\href {\doibase 10.1109/MCSE.2011.37} {\bibfield  {journal} {\bibinfo  {journal} {Comput. Sci. Eng.}\ }\textbf {\bibinfo {volume} {13}},\ \bibinfo {pages} {22} (\bibinfo {year} {2011})},\ \Eprint {http://arxiv.org/abs/1102.1523} {arXiv:1102.1523 [cs.MS]} \BibitemShut {NoStop}%
\bibitem [{\citenamefont {Virtanen}\ \emph {et~al.}(2020)\citenamefont {Virtanen} \emph {et~al.}}]{Virtanen:2019joe}%
  \BibitemOpen
  \bibfield  {author} {\bibinfo {author} {\bibfnamefont {P.}~\bibnamefont {Virtanen}} \emph {et~al.},\ }\href {\doibase 10.1038/s41592-019-0686-2} {\bibfield  {journal} {\bibinfo  {journal} {Nature Meth.}\ }\textbf {\bibinfo {volume} {17}},\ \bibinfo {pages} {261} (\bibinfo {year} {2020})},\ \Eprint {http://arxiv.org/abs/1907.10121} {arXiv:1907.10121 [cs.MS]} \BibitemShut {NoStop}%
\bibitem [{\citenamefont {Hunter}(2007)}]{Hunter:2007ouj}%
  \BibitemOpen
  \bibfield  {author} {\bibinfo {author} {\bibfnamefont {J.~D.}\ \bibnamefont {Hunter}},\ }\href {\doibase 10.1109/MCSE.2007.55} {\bibfield  {journal} {\bibinfo  {journal} {Comput. Sci. Eng.}\ }\textbf {\bibinfo {volume} {9}},\ \bibinfo {pages} {90} (\bibinfo {year} {2007})}\BibitemShut {NoStop}%
\bibitem [{\citenamefont {Foreman-Mackey}(2016)}]{corner}%
  \BibitemOpen
  \bibfield  {author} {\bibinfo {author} {\bibfnamefont {D.}~\bibnamefont {Foreman-Mackey}},\ }\href {\doibase 10.21105/joss.00024} {\bibfield  {journal} {\bibinfo  {journal} {The Journal of Open Source Software}\ }\textbf {\bibinfo {volume} {1}},\ \bibinfo {pages} {24} (\bibinfo {year} {2016})}\BibitemShut {NoStop}%
\end{thebibliography}%
\end{document}